\documentclass[12pt]{article}
\usepackage{amssymb,amsmath,bm,bbm}
\usepackage{epsf}
\usepackage{epsfig}
\usepackage{afterpage}
\usepackage{longtable}
\usepackage{cite}
\usepackage{latexsym, mathrsfs}
\usepackage{axodraw}
\usepackage{graphics}
\usepackage{color}

\newcommand{\rep}[1]{\mathbf{#1}}

\newcommand{\singlet}{(\rep{1},\rep{1})}

\newcommand{\Ltriplet}{(\rep{3},\rep{1})}
\newcommand{\Rtriplet}{(\rep{1},\rep{3})}

\def\slash#1{\setbox0=\hbox{$#1$}\dimen0=\wd0
      \setbox1=\hbox{/} \dimen1=\wd1 \ifdim\dimen0>\dimen1
      \rlap{\hbox to \dimen0{\hfil/\hfil}} #1                        \else
      \rlap{\hbox to \dimen1{\hfil$#1$\hfil}}
      /   \fi}

\newcommand{\lsim}{
\mathrel{\hbox{\rlap{\hbox{\lower4pt\hbox{$\sim$}}}\hbox{$<$}}}}

\newcommand{\gsim}{
\mathrel{\hbox{\rlap{\hbox{\lower4pt\hbox{$\sim$}}}\hbox{$>$}}}}

\newcommand{\vcb}{|V_{cb}|}
\newcommand{\vtd}{|V_{td}|}

\newcommand{\vts}{|V_{ts}|}

\def\eps{\varepsilon}

\newcommand{\tev}{\, {\rm TeV}}
\newcommand{\gev}{\, {\rm GeV}}
\newcommand{\mev}{\, {\rm MeV}}

\newcommand{\Heff}{{\cal H}_\text{eff}}

\def\beq{\begin{equation}}
\def\eeq{\end{equation}}

\newcommand{\be}{\begin{equation}}
\newcommand{\ee}{\end{equation}}
\newcommand{\bea}{\begin{eqnarray}}
\newcommand{\eea}{\end{eqnarray}}
\newcommand{\nn}{\nonumber}
\newcommand{\bi}{\begin{itemize}}
\newcommand{\ei}{\end{itemize}}
\newcommand{\ord}{{\cal O}}

\def\kpn{K^+\rightarrow\pi^+\nu\bar\nu}

\def\klpn{K_{L}\rightarrow\pi^0\nu\bar\nu}

\newcommand{\newsection}[1]{\section{#1}\setcounter{equation}{0}}

\begin{document}
\begin{titlepage}
\vspace*{-0.5truecm}

\begin{flushright}
{TUM-HEP-704/08}\\
{MPP-2008-152}
\end{flushright}

\vfill

\begin{center}
\boldmath
{\Large\textbf{Rare $K$ and $B$ Decays in a \vspace{1mm}\\Warped Extra Dimension with Custodial Protection}}
\unboldmath
\end{center}

\begin{center}
{\bf  Monika Blanke$^{a,b}$, Andrzej J.~Buras$^{a,c}$, Bj\"orn Duling$^a$,\\  Katrin Gemmler$^a$ and Stefania Gori$^{a,b}$}
\vspace{0.3truecm}

{\footnotesize
 $^a${\sl Physik Department, Technische Universit\"at M\"unchen,
D-85748 Garching, Germany}\vspace{0.2cm}

 {\sl $^b$Max-Planck-Institut f{\"u}r Physik (Werner-Heisenberg-Institut), \\
D-80805 M{\"u}nchen, Germany}\vspace{0.2cm}

 $^c${\sl TUM Institute for Advanced Study, Technische Universit\"at M\"unchen, \vspace{-0.1cm}\\ D-80333 M\"unchen, Germany}\vspace{0.2cm}
}
\end{center}

\begin{abstract}
{\small\noindent
We present a complete study of rare $K$ and $B$ meson decays 
in a warped extra dimensional model with a custodial protection 
of (both diagonal and non-diagonal) $Z d_L^i \bar d_L^j$ couplings,
including $K^+\to \pi^+\nu\bar\nu$, $K_L\to\pi^0\nu\bar\nu$, $K_L\to\pi^0 \ell^+\ell^-$,
$K_L\to \mu^+\mu^-$, $B_{s,d}\to \mu^+\mu^-$, $B\to K\nu\bar\nu$, 
$B\to K^*\nu\bar\nu$  and $B\to X_{s,d}\nu\bar\nu$.
 In this model in addition to Standard Model one loop contributions
 these processes receive  tree level contributions from the $Z$ boson and the new heavy electroweak gauge bosons. We analyse all these contributions that turn out to be dominated by tree level $Z$ boson exchanges governed by right-handed couplings to down-type quarks. Imposing
all existing constraints from $\Delta F=2$ transitions analysed by us
recently and fitting all quark masses and CKM mixing parameters we 
find that a number of branching ratios for rare $K$ decays 
can differ significantly from the SM predictions, while the corresponding effects in rare $B$ decays are modest, dominantly due to the custodial protection being more effective in
 $B$ decays than in $K$ decays. In order to reduce the parameter dependence we study correlations between various observables within the $K$ system, within the $B$ system and in 
particular between $K$ and $B$ systems, and also between $\Delta F=2$ and $\Delta F=1$ observables. These correlations allow for a clear
distinction between this new physics scenario and models with
minimal flavour violation or the Littlest Higgs Model with T-parity, and could give an opportunity to future experiments to confirm or rule out the model.
  We  show how our results would change if the custodial protection of 
$Z d_L^i \bar d^j_L$ couplings was absent. In the case of rare $B$ decays the modifications are spectacular.}
\end{abstract}

%
%
%
\end{titlepage}

\setcounter{page}{1}
\pagenumbering{arabic}

\newsection{Introduction}\label{sec:int}

In a recent paper \cite{Blanke:2008zb} we have presented a complete study of 
$\Delta S=2$ and $\Delta B=2$ processes in a {Randall-Sundrum (RS) \cite{Randall:1999ee} model}
with an extended gauge group and the custodial $P_{LR}$ symmetry that has 
been constructed
to protect the $T$ parameter and the coupling $Zb_L\bar b_L $ from new physics {(NP)}
tree level contributions \cite{Agashe:2003zs,Csaki:2003zu,Agashe:2006at}. In this context we have pointed out  {\cite{Blanke:2008zb}} that this
 custodial symmetry {automatically protects} flavour violating
 $Zd_L^i\bar d_L^j$ couplings so that tree level $ Z$ contributions to 
 all processes in which the flavour changes appear in the down quark sector
are  dominantly represented by $Zd_R^i\bar d_R^j$ couplings.\footnote{The $Z$ tree level contribution to $\Delta F=2$ 
processes is $\ord{(v^4/M_{\text{KK}}^4})$ and can be neglected.}

{Additional { NP} contributions to the decays in question in the RS model 
considered arise} from tree level heavy electroweak gauge bosons $Z_H$ and $Z^\prime$
and KK photon $A^{(1)}$
exchanges.
{However} the $Z^\prime d_L^i\bar d_L^j$ couplings are suppressed, similarly 
to $Z d_L^i\bar d_L^j$ couplings, by the custodial symmetry, and $A^{(1)}$
contributions are suppressed by the electromagnetic coupling constant $e^\text{4D}$ and the electric charge $Q=-1/3$ of down-type quarks. Consequently only 
$Z_H$ 
contributions are really relevant.
They  played  
a prominent role in
$\Delta B=2$ observables considered in \cite{Blanke:2008zb} but had to compete
 there
with the KK gluon exchanges. The latter contributions are 
absent at tree level in  $\Delta F=1$ decays with leptons in the final state
 and consequently at first sight one would expect that rare $K$ and 
$B$ decays are governed by the physics of the
$Z_H$  gauge boson. However, in $\Delta F=1$ processes 
tree level $Z$ contributions are of the same order in $ v^2/M_{\text{KK}}^2$ as
the contribution from $Z_H$, and moreover $Z$ boson
couplings to leptons are $\ord(1)$, whereas the ones of $Z_H$ and $Z^\prime$
are suppressed. Consequently, $Z_H$ has to
compete this time with tree level $Z$ contributions, and as we will
see below $Z$ generally wins this competition in spite of the
{custodial} protection of its left-handed couplings to the {down-type} quarks.
All these new effects 
bring in new flavour violating 
interactions beyond those governed by the CKM matrix and one should expect
an interesting pattern of deviations from the SM {and minimal flavour violation (MFV) \cite{Buras:2000dm,Buras:2003jf,D'Ambrosio:2002ex,Chivukula:1987py,Hall:1990ac}} expectations.

The goal of the present paper is to extend our analysis of $\Delta F=2$
processes in \cite{Blanke:2008zb} to rare  decays of $K$ and $B_{d,s}$ mesons.
In particular we will present for the first time the formulae for the
branching ratios  of
  $K^+\to \pi^+\nu\bar\nu$, $K_L\to\pi^0\nu\bar\nu$, $K_L\to\pi^0 \ell^+\ell^-$,
$K_L\to \mu^+\mu^-$, $B_{s,d}\to \mu^+\mu^-$, $B\to K\nu\bar\nu$, 
$B\to K^*\nu\bar\nu$  and
$B_{s,d}\to X_{s,d}\nu\bar\nu$
  in the warped extra dimensional  model with a protective custodial symmetry.
A partial study of these decays  in a model without {custodial} {protection} has been done in {\cite{Burdman:2002gr,Burdman:2003nt,Agashe:2004cp,Moreau:2006np}} {and a more detailed analysis in the latter case is in progress \cite{Haisch-BF08}}.

Two comments should be made already at the beginning of our paper:
\begin{itemize}
\item
It is known that the model in question cannot easily satisfy the 
$\varepsilon_K$ constraint for KK scales of order {$\ord(1\tev)$} \cite{Csaki:2008zd}.\footnote{The same conclusion has been reached in the two-site approach in \cite{Agashe:2008uz}.} Yet
as we have demonstrated in \cite{Blanke:2008zb} 
there exist regions in parameter space with only modest fine-tuning
in the 5D Yukawa couplings involved, which {allows} to obtain a satisfactory
description of the quark masses and {CKM parameters} and to satisfy all
existing $\Delta F=2$ {(in particular $\eps_K$)} and
electroweak precision constraints for scales $M_\text{KK}\simeq 3\tev$
in {the} reach of the LHC. In the present paper we {will perform}
 our numerical
analysis for these regions of parameter  space {only}.
\item
In view of many free parameters in the model in question we will search
for correlations between various observables with the hope that these
correlations will be less parameter dependent than the individual observables
themselves. Such correlations can originate {from} the fact that the quark
shape functions enter various processes universally. 
\end{itemize}

Our paper is organised as follows. In Section \ref{sec:mod} we summarise briefly 
those ingredients of the model in question that we will need in our analysis.
A very detailed presentation of the model is presented in 
\cite{Blanke:2008aa}. 
In Section \ref{sec:Heff} we derive the formulae for the effective
Hamiltonians governing $s\to d\nu\bar\nu$, $b\to q \nu\bar\nu$, $s\to
d\ell^+\ell^-$ and $b\to q \ell^+\ell^-$ ($q=d,s$) transitions.
In Section \ref{sec:excl} we calculate  {the}
most interesting exclusive rare decay
 branching ratios in {the} $K$ and $B$ meson systems, including {those for the processes} $\kpn$,
$\klpn$, $B\to K^{(*)}\nu\bar\nu$, $B_{s,d}\to\mu^+\mu^-$, $K_L \to\mu^+\mu^-$
 and $K_L \to \pi^0\ell^+\ell^-$. Section \ref{sec:incl} is dedicated to the
 inclusive decays $B\to X_{s,d}\nu\bar\nu$. In Section \ref{sec:correlations}
we outline the strategy for the study of {a number of} correlations
 between 
various observables within the $K$ system, within the $B$ system and in 
particular between $K$ and $B$ systems.
{Sections \ref{sec:Heff}--\ref{sec:correlations}}
give formulae  
that are sufficiently general to
be applied to {every} model with tree level flavour violating contributions in
which heavy neutral gauge bosons have arbitrary masses and arbitrary
left-handed and right-handed couplings. Moreover several ideas, in particular
related to correlations between various observables are applicable to 
all extensions of the SM.
In Section \ref{sec:anatomy}, before entering the numerical analysis,
 we present the anatomy of NP
contributions in the model in question
 that reveals a particular pattern of deviations from the SM. 
In Section \ref{sec:num} a detailed
numerical analysis of these branching ratios is presented.
In particular we study the correlations not only between various $\Delta F=1$
observables but also between $\Delta F=1$ and $\Delta F=2$ observables.
Of interest is the question whether the large values of
 $A_\text{CP}(B_s\to\psi\phi)$ and $A^s_\text{SL}$ found in \cite{Blanke:2008zb}
can still be found simultaneously with large effects in rare decay 
branching ratios.
 We summarise
our results in Section \ref{sec:summary}. Few technicalities are relegated to appendices.

\newsection{The Model}\label{sec:mod}

\subsection{Preliminaries}

The aim of this section is to briefly review the most important ingredients of the model under consideration. A detailed theoretical discussion is presented in \cite{Blanke:2008aa}. 

The starting point is the Randall-Sundrum (RS) geometric background, i.\,e.\;a 5D spacetime, where the extra dimension is {restricted} to the interval $0\le y\le L$, with a warped metric given by
\cite{Randall:1999ee}
\be\label{eq:RS}
ds^2=e^{-2ky}\eta_{\mu\nu}dx^\mu dx^\nu - dy^2\,.
\ee
Here the curvature scale $k$ is assumed to be $k\sim\ord(M_\text{Pl})$. Due to the exponential warp factor $e^{-ky}$, the effective energy scales depend on the position $y$ along the extra dimension, so that with 
$e^{kL}=10^{16}$ the gauge hierarchy problem can be solved. In what follows  we will therefore treat
\be\label{eq:f}
f=ke^{-kL}\sim \ord(\text{TeV})
\ee
as the only free parameter coming from space-time geometry. {In our numerical analysis we will use $f=1\tev$.}

In order to avoid stringent constraints from electroweak precision observables, all gauge and matter fields are assumed to live in the 5D bulk \cite{Gherghetta:2000qt,Chang:1999nh,Grossman:1999ra}, while the Higgs boson is confined to the IR brane at $y=L$.

\subsection{Electroweak Gauge Sector}

The minimal RS model with bulk {fields} and the SM gauge group in the bulk turns out to be severely constrained by EW precision data and in particular {by} the $T$ parameter, so that the first KK excitations have to be as heavy as $M_\text{KK}\sim\ord(10\tev)$ and {would consequently be} beyond the reach of the LHC \cite{Agashe:2003zs,Casagrande:2008hr}.

However such dangerous contribution to the $T$ parameter and also to the $Zb_L\bar b_L$ coupling can be avoided by enlarging the bulk {symmetry} to \cite{Agashe:2003zs,Csaki:2003zu,Agashe:2006at}
\be
G_\text{bulk}=SU(3)_c\times SU(2)_L\times SU(2)_R\times U(1)_X\times P_{LR}\,.
\ee
Here the discrete exchange symmetry 
\be
P_{LR}: SU(2)_L \leftrightarrow SU(2)_R
\ee
has been introduced in order to suppress the non-universal corrections to the
$Zb_L\bar b_L$ vertex. Details on {particular fermion embeddings} respecting
that symmetry can be found  e.\,g.~in 
{\cite{Agashe:2006at,Carena:2006bn,Cacciapaglia:2006gp,Blanke:2008aa}}.

From the enlarged gauge group and the first excited KK modes of the SM 
electroweak gauge bosons that we include in our analysis
 there arise three new neutral electroweak gauge bosons,
\be
Z_H\,,\qquad Z'\,, \qquad A^{(1)}
\ee
in addition to the SM $Z$ boson and photon, where the first two are linear
combinations of the gauge eigenstates $Z^{(1)}$ and $Z_X^{(1)}$
\cite{Blanke:2008aa}. Neglecting small $SU(2)_R$ breaking effects on the UV
brane ($y=0$) and corrections due to EW symmetry breaking, one finds\footnote{
We would like to caution the reader that a different notation has been used in \cite{Casagrande:2008hr}: Their $M_\text{KK}$ corresponds to our $f$, so that  in spite of comparable $M_\text{KK}$  the masses of
the first KK gauge bosons in that paper are  larger  than in our analysis.}
\be\label{degeneracymasses}
M_{Z_H}=M_{Z'}=M_{A^{(1)}} \equiv M_\text{KK} \simeq 2.45f\,.
\ee 
All KK gauge bosons are localised close to the IR brane, inducing tree level FCNCs, as discussed in more detail in Section \ref{sec:flavour}. 

The RS model with custodial protection of \,$T$ and $Zb_L\bar b_L$ as described above {and mildly constrained quark shape functions} then turns out to be consistent with EW precision observables for KK scales as low as $M_\text{KK}\simeq(2-3)\tev$ \cite{Contino:2006qr,Carena:2007ua}.

\subsection{Fermion Sector}

\subsubsection{Zero Mode Localisation}

Bulk fermions in the RS background offer a natural explanation of the observed hierarchies in fermion masses and mixings \cite{Grossman:1999ra,Gherghetta:2000qt,Huber:2003tu}. At the same time a powerful suppression mechanism for flavour changing neutral current (FCNC) interactions, the so-called \emph{RS-GIM mechanism}, is provided \cite{Agashe:2004cp}.

The bulk profile of a fermionic zero mode depends strongly on its bulk mass parameter $c_\Psi$. In case of a left-handed zero mode $\Psi_L^{(0)}$ it is given by
\be
f^{(0)}_L(y,c_\Psi) = \sqrt{\frac{(1-2c_\Psi)kL}{e^{(1-2c_\Psi)kL}-1}} e^{- c_\Psi ky}
\ee
with respect to the warped metric. Therefore, for $c_\Psi >1/2$ the fermion $\Psi_L^{(0)}$ is localised towards the UV brane and exponentially suppressed on the IR brane, while for $c_\Psi < 1/2$ it is localised towards the IR brane. The bulk profile for a right-handed zero mode $\Psi_R^{(0)}$ can be obtained from
\be
f^{(0)}_R(y,c_\Psi) = f^{(0)}_L(y,-c_\Psi)\,,
\ee
so that its localisation depends on whether $c_\Psi <-1/2$ or $c_\Psi >-1/2$. Note that as in the SM the left- and right-handed zero modes present in the spectrum necessarily belong to different 5D multiplets, so that generally $c_{\Psi_L} \ne c_{\Psi_R}$.

In order to preserve the discrete $P_{LR}$ symmetry, we embed the left handed SM quarks into bidoublets of $SU(2)_L\times SU(2)_R$, while the right handed up and down quarks belong to $\singlet$ and {$\Rtriplet\oplus\Ltriplet$} representations, respectively \cite{Carena:2006bn,Blanke:2008aa}. Their bulk mass parameters are denoted by 
 $c_Q^i$, $c_{u}^i$ and $c_{d}^i$, respectively.

\subsubsection{Higgs Field and Yukawa Couplings}\label{sec:2.4}

As the Higgs boson in our model is localised on the IR brane, the effective 4D Yukawa couplings, relevant for the SM fermion masses and mixings, read
\be\label{eq:Yud}
Y^{u,d}_{ij} = \lambda^{u,d}_{ij}\,\frac{e^{kL}}{kL} f^{(0)}_{L}(y=L,c_Q^i)  f^{(0)}_{R}(y=L,c_{u,d}^j) \equiv \lambda^{u,d}_{ij}\,\frac{e^{kL}}{kL} f^Q_i  f^{u,d}_j \,,
\ee
where $\lambda^{u,d}$ are the fundamental 5D Yukawa coupling matrices. In order to preserve perturbativity of the model, we require as usual $|\lambda^{u,d}_{ij}|\le3$. Here and in the following we work in the special basis in which the bulk mass matrices are taken to be real and diagonal. Such a basis can always be reached by appropriate unitary transformations in the $Q_i$, $u_i$ and $d_i$ sectors.

Note that the strong hierarchies of quark masses and mixings can be traced back to $\ord(1)$ bulk {mass parameters} and anarchic 5D Yukawa couplings  $\lambda^{u,d}$ due to the exponential dependence of $Y^{u,d}$ on the bulk mass parameters $c_{Q,u,d}$. The flavour structure then resembles the one of models with a Froggatt-Nielsen symmetry \cite{Froggatt:1978nt}, so that with the help of the latter paper analytic formulae for quark masses and flavour mixing matrices $\mathcal{U}_L,\mathcal{U}_R,\mathcal{D}_L,\mathcal{D}_R$ and the CKM matrix
\be
V_\text{CKM}= \mathcal{U}_L^\dagger\mathcal{D}_L
\ee
can be obtained \cite{Blanke:2008zb,Casagrande:2008hr}.

Due to the mixing with heavy KK fermions flavour violating Higgs couplings are 
induced already at tree level. However it can straightforwardly be shown 
\cite{Blanke:2008zb} that these couplings receive a strong chirality 
suppression in addition to the usual RS-GIM suppression and are therefore 
negligibly small.

\subsubsection{Flavour Violation by Neutral Electroweak Gauge Bosons}\label{sec:flavour}

As a consequence of the exponential localisation of the gauge KK modes
towards the IR brane, their couplings to zero mode fermions are not flavour universal, but depend strongly on the relevant bulk mass parameter $c_{Q,u,d}^i$. After rotation to the fermion mass eigenbasis then FCNC couplings of the heavy gauge bosons $Z_H$, $Z'$ and $A^{(1)}$ are induced. They can be parameterised by the coupling matrices $\hat \Delta_{L}(V)$ and $\hat \Delta_{R}(V)$ ($V=Z_H,Z',A^{(1)}$), that have been evaluated in \cite{Blanke:2008zb} and are collected for completeness in Appendix \ref{app:Deltas}.

{Moreover}, due to the mixing of the SM $Z$ boson with the heavy KK modes $Z^{(1)}$ and $Z^{(1)}_X$, also the $Z$ boson couplings become flavour violating already at the tree level. {An additional contribution arises from the mixing of the zero mode fermions with their heavy KK partners.} 

On the other hand, it has been observed in \cite{Blanke:2008zb} that the custodial protection symmetry $P_{LR}$, originally introduced to protect the $Zb_L\bar b_L$ coupling, removes not only the non-universal contributions to that coupling, but at the same time efficiently protects all tree level $Z d^i_L\bar d^j_L$ couplings, so that at the tree level the $Z$ boson couplings to left-handed down type quarks are strongly suppressed. {We note that the protective $P_{LR}$ symmetry is at work not only for the interplay of $Z^{(1)}$ and $Z^{(1)}_X$ contributions to the {$Z$ and $Z'$} couplings, but also for the KK fermion contributions. {This is} because {the 
fermion sector has been constructed in a $P_{LR}$-invariant manner as well and the left-handed down-type quarks transform as $P_{LR}$-eigenstates.}}
The protective symmetry is however not active for right-handed  quarks, so that flavour violating $Z d^i_R\bar d^j_R$ couplings {are important} already at the tree level. 

While $Z$ boson contributions to $\Delta F=1$ decays are parametrically enhanced with respect to the contributions of {$Z_H$, $Z'$ and $A^{(1)}$} by a factor $kL\sim 35$ \cite{Agashe:2004cp,Casagrande:2008hr}, they are {suppressed by} the fact that flavour violation is generally weaker in the right-handed sector.
In spite of this we will see in Sections \ref{sec:anatomy} and \ref{sec:num} 
that $Z$ boson contributions are larger than the contributions of {$Z_H,Z'$ and $A^{(1)}$}.

We note that the strength of RS flavour violation depends on the presence of 
possible brane kinetic terms which alter the matching relation between 5D and 
4D gauge couplings, see \cite{Csaki:2008zd,Csaki:2008eh,Blanke:2008zb} for 
details. In order not to complicate the present analysis we assume the simple 
intermediate scenario in which the tree level matching condition $g^\text{4D} 
=g^\text{5D}/\sqrt{L}$ holds. A generalisation of our analysis to include 
deviations from this ansatz is straightforward.

\newsection{Basic Formulae for Effective Hamiltonians}\label{sec:Heff}

\subsection{Preliminaries}
The goal of the present section is to give formulae for the effective
Hamiltonians relevant for rare $K$ and $B$ decays that in addition to
SM one-loop contributions include tree level contributions from the SM
$Z$ gauge boson and
{the heavy gauge bosons $Z_H$, $Z^\prime$ and $A^{(1)}$}.
It will be useful to {first keep} our presentation as general as possible so
that the formulae given below could be applied to {all} other models
with tree level  flavour violating contributions in which the 
 heavy neutral gauge bosons
have arbitrary masses and arbitrary left-handed and right-handed couplings.
Subsequently, we will apply these formulae to our model in which at
leading order the three new heavy electroweak gauge bosons {are degenerate} in mass and certain simplifications
{occur}.

Our presentation includes the contributions from all operators originating
only from tree level exchanges of electroweak gauge bosons. Consequently we
do not discuss the dipole operators that enter effective Hamiltonians first
at the one-loop level. We will return to them in the context of the model
in question in a separate publication. This implies that the effective
Hamiltonians for $b\to d \ell^+\ell^-$ and $b\to s \ell^+\ell^-$ transitions given below 
are 
incomplete and we cannot perform yet the phenomenology of decays like
$B\to K^*\ell^+\ell^-$, $B\to X_{s,d} \ell^+\ell^-$ and $B\to X_{s,d}\gamma$ that is
left for the future.

\boldmath
\subsection{Effective Hamiltonian for $s\to d\nu\bar\nu$}\label{sec:sdnn}
\unboldmath

The effective Hamiltonian for $s \rightarrow d\nu\bar\nu$ transitions is given in the SM as follows
\be\label{Heffpr}
\left[\Heff^{\nu\bar\nu}\right]^K_\text{SM}=g_{\text{SM}}^2\sum_{\ell=e,\mu,\tau}{\left[\lambda_c^{(K)} {X_\text{NNL}^\ell(x_c)}+\lambda_t^{(K)} X(x_t)\right]}(\bar sd)_{V-A}(\bar\nu_\ell\nu_\ell)_{V-A}+h.c.\,,
\ee
where $x_i=m_i^2/M_W^2$, $\lambda_i^{(K)} =V_{is}^*V_{id}^{}$ and $V_{ij}$ are the elements of the CKM matrix. {$X_\text{NNL}^\ell(x_c)$} and $X(x_t)$ comprise 
internal charm and top quark contributions, respectively. They are known to high accuracy including QCD corrections \cite{Buchalla:1998ba,Buras:2005gr,Buras:2006gb}. For convenience we have introduced
\be\label{gsm}
g_{\text{SM}}^2=\frac{G_F}{\sqrt 2}\frac{\alpha}{2\pi\sin^2\theta_W}\,.
\ee

In the RS model considered, $\left[\Heff^{\nu\bar\nu}\right]^K$ receives
tree-level contributions from $Z$ and from the heavy neutral gauge bosons $Z_H$ and $Z'$
which contain new flavour violating interactions.

We begin with the FCNC Lagrangian for $Z$
\be\label{eq:LeffZ1}
\mathcal{L}_\text{FCNC}(Z)=-\left[\mathcal L_L(Z)+\mathcal L_R(Z)\right]\,,
\ee
where
\begin{eqnarray}
\mathcal L_L(Z)&=& \Delta_L^{sd}(Z)\left(\bar s_L\gamma_\mu d_L\right)Z^\mu\,,\\
\mathcal L_R(Z)&=& \Delta_R^{sd}(Z)\left(\bar s_R\gamma_\mu d_R\right)Z^\mu\,,
\end{eqnarray}
with explicit expressions for $\Delta_L^{sd}(Z)$ and $\Delta_R^{sd}(Z)$ given in Appendix \ref{app:Deltas}. 

For the $Z\nu\bar\nu$ coupling we analogously write
\be\label{Lnunu}
\mathcal L_{\nu\bar\nu}(Z)=  -\Delta_L^{\nu\nu}(Z)(\bar\nu_L\gamma_\mu\nu_L)Z^\mu
\ee
where $\Delta_L^{\nu\nu}(Z)$ is given in Appendix \ref{app:Deltas}.

\begin{figure}
\begin{center}
\begin{picture}(100,100)(0,0)
\ArrowLine(0,10)(30,40)
\ArrowLine(30,40)(0,70)
\Photon(30,40)(100,40){2}{6}
\Vertex(30,40){1.3}
\Vertex(100,40){1.3}
\ArrowLine(100,40)(130,70)
\ArrowLine(130,10)(100,40)
\Text(65,21)[cb]{{\Black{$Z,Z^\prime,Z_H$}}}
\Text(20,60)[cb]{{\Black{$s$}}}
\Text(20,13)[cb]{{\Black{$d$}}}
\Text(110,60)[cb]{{\Black{$\nu$}}}
\Text(110,13)[cb]{{\Black{$\nu$}}}
\end{picture}
\end{center}
\caption{\it Tree level contributions of $Z$, $Z^\prime$ and $Z_H$ to the $s\to d\nu\bar\nu$ effective Hamiltonian.\label{KplusZ_H}}
\end{figure}
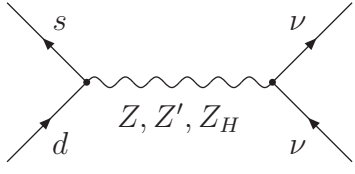

A straightforward calculation of the diagram in Fig.\ \ref{KplusZ_H} results in a new contribution to $\left[\Heff^{\nu\bar\nu}\right]^K$
\be
\left[\Heff^{\nu\bar\nu}\right]_{Z}^K=\frac{\Delta_L^{\nu\nu}(Z)}{M_{Z}^2}
\left[\Delta_L^{sd}(Z)(\bar s_L\gamma^\mu d_L)+\Delta_R^{sd}(Z)(\bar s_R\gamma^\mu d_R)\right]
\left(\bar\nu_L\gamma_\mu\nu_L\right)+h.c.\,.\label{eq:HeffKnunuZ1}
\ee

The contribution of $Z^\prime$ and $Z_H$ to $\left[\Heff^{\nu\bar\nu}\right]^K$ can then be obtained from \eqref{eq:HeffKnunuZ1} by simply replacing $Z$ by $Z^\prime$ and $Z_H$ respectively. Explicit expressions for $\Delta_{L,R}^{sd}(Z^\prime)$, $\Delta_L^{\nu\nu}(Z^\prime)$ and $\Delta_{L,R}^{sd}(Z_H)$, $\Delta_L^{\nu\nu}(Z_H)$ can be found in Appendix \ref{app:Deltas}.

Combining then the contributions of $Z, Z^\prime, Z_H$ in \eqref{eq:HeffKnunuZ1} with the SM contribution in  \eqref{Heffpr},
\be
\left[\Heff^{\nu\bar\nu}\right]^K = \left[\Heff^{\nu\bar\nu}\right]^K_\text{SM}+
\left[\Heff^{\nu\bar\nu}\right]_{Z}^K+\left[\Heff^{\nu\bar\nu}\right]_{Z^\prime}^K+\left[\Heff^{\nu\bar\nu}\right]_{Z_H}^K\,,
\ee
we finally find
\bea
\left[\Heff^{\nu\bar\nu}\right]^K &=&
g_{\text{SM}}^2\sum_{\ell=e,\mu,\tau}
{\left[\lambda_c^{(K)} {X_\text{NNL}^\ell(x_c)}+\lambda_t^{(K)}X_{K}^{V-A}\right]}
(\bar sd)_{V-A}(\bar\nu_\ell\nu_\ell)_{V-A}\nn\\
&& {}
+ g_{\text{SM}}^2\sum_{\ell=e,\mu,\tau}{\left[\lambda_t^{(K)}X_K^V\right]}(\bar sd)_{V}(\bar\nu_\ell\nu_\ell)_{V-A}
+h.c.\,.
\eea
Here we have introduced the functions $X_K^{V-A}$ and $X_K^V$, generalising
the structure encountered in the {Littlest Higgs model with T-parity (LHT)} in 
\cite{Blanke:2006eb},
\bea
X_K^{V-A} &=& X(x_t) + \sum_{i=Z,Z^\prime,Z_H} (X_{i}^K)^{V-A}\,,\\
 X_K^V &=& \sum_{i=Z,Z^\prime,Z_H} (X_{i}^K)^{V} \,,
\eea
that will turn out to be useful later on. The $Z$ contributions are given as follows
\bea
(X_{Z}^K)^{V-A} &=& \frac{1}{\lambda_t^{(K)}} \frac{\Delta_L^{\nu\nu}(Z)}{4M_{Z}^2 g_\text{SM}^2} \left[\Delta_L^{sd}(Z)-\Delta_R^{sd}(Z)\right]\,,\label{eq:XKZ1}\\
(X_{Z}^K)^V &=& \frac{1}{\lambda_t^{(K)}} \frac{\Delta_L^{\nu\nu}(Z)}{2 M_{Z}^2 g_\text{SM}^2} \Delta_R^{sd}(Z)\,,\label{eq:tildeXKZ1}
\eea
and the $Z^\prime$, $Z_H$ contributions can be obtained from \eqref{eq:XKZ1} and \eqref{eq:tildeXKZ1} by simply replacing $Z$ by $Z^\prime$ and $Z_H$ respectively.\footnote{Note that the expression for $g_\text{SM}$ is not modified and remains as defined in \eqref{gsm}.}

Some comments are in order:
\bi
\item
In the SM only a single operator $(\bar sd)_{V-A}(\bar\nu\nu)_{V-A}$ is present. This is due to the purely left-handed structure of $SU(2)_L$ gauge couplings and therefore of FCNC transitions.
\item
In the RS model in question also the operator $(\bar sd)_{V}(\bar\nu\nu)_{V-A}$ is present, as both the {$\hat\Delta_L$ and $\hat\Delta_R$} coupling matrices have non-diagonal entries. {Indeed it will turn out that in most cases $\Delta_R^{sd}(Z)$ yields the dominant contribution.}
\item
On the other hand in the RS model under consideration the gauge couplings of the neutrino zero modes are purely left-handed, as the right-handed neutrinos are introduced as gauge singlets \cite{Blanke:2008aa}.
\item
{As all NP contributions have been collected in the term proportional to $\lambda_t^{(K)}$, {$X_\text{NNL}^\ell(x_c)$} contains only the SM contributions} that are known including QCD corrections at the NNLO level \cite{Buras:2005gr,Buras:2006gb}.
\ei

\boldmath
\subsection{Effective Hamiltonian for $b\to d\nu\bar\nu$ and $b\to s\nu\bar\nu$}
\label{sec:bqnn}\unboldmath

Let us now generalise the result obtained in the previous section to the case of $b\to d\nu\bar\nu$ and $b\to s\nu\bar\nu$ transitions. Basically only two steps have to be performed:
\begin{enumerate}
\item
All flavour indices have to be adjusted appropriately.
\item
The charm quark contribution can be safely neglected in $B$ physics, so that
only
\be
\lambda_t^{(d)} =V_{tb}^*V_{td}, \qquad \lambda_t^{(s)} =V_{tb}^*V_{ts}
\ee
enter the expressions below.
\end{enumerate}
The effective Hamiltonian for $b\to q\nu\bar\nu$ ($q=d,s$) is then given as follows:
\bea
\left[\Heff^{\nu\bar\nu}\right]^{B_q} &=&
g_{\text{SM}}^2\sum_{\ell=e,\mu,\tau}
{\left[\lambda_t^{(q)}X_q^{V-A}\right]}
(\bar bq)_{V-A}(\bar\nu_\ell\nu_\ell)_{V-A}\nn\\
&& + g_{\text{SM}}^2\sum_{\ell=e,\mu,\tau}{\left[\lambda_t^{(q)} X_q^V\right]}(\bar bq)_{V}(\bar\nu_\ell\nu_\ell)_{V-A}
+h.c.\,,
\eea
with
\bea
X_q^{V-A} &=& X(x_t) +   \sum_{i=Z,Z^\prime,Z_H} (X_{i}^q)^{V-A}
\,,\\
X_q^V &=&   \sum_{i=Z,Z^\prime,Z_H} (X_{i}^q)^{V}
\eea
and
\bea
(X_{Z}^q)^{V-A} &=& \frac{1}{\lambda_t^{(q)}} \frac{\Delta_L^{\nu\nu}(Z)}{4M_{Z}^2 g_\text{SM}^2} \left[\Delta_L^{bq}(Z)-\Delta_R^{bq}(Z)\right]\,,\label{eq:XqZ1}\\
(X_{Z}^q)^V &=& \frac{1}{\lambda_t^{(q)}} \frac{\Delta_L^{\nu\nu}(Z)}{2 M_{Z}^2 g_\text{SM}^2} \Delta_R^{bq}(Z)\,.\label{eq:tildeXqZ1}
\eea
The $Z^\prime$ and $Z_H$ contributions can be obtained from \eqref{eq:XqZ1} and \eqref{eq:tildeXqZ1} by simply replacing $Z$ by $Z^\prime$ and $Z_H$ respectively. Again all relevant {$\Delta_{L,R}^{bq}$ entries} can be found in Appendix~\ref{app:Deltas}. 

Note that the functions $X_{K,d,s}^{V-A,V}$ depend on the quark flavours involved, through the flavour indices in the $\Delta_{L,R}^{ij}$ {$(i,j=s,d,b)$} couplings and through the $1/\lambda_t^{(q)}$ ($q=K,d,s$) factor in front of the new RS contributions. 
 This should be contrasted {with} the case of the SM where $K$, $B_d$ and $B_s$ systems are governed by a \emph{flavour-universal} loop function $X(x_t)$ and the only flavour dependence enters through the CKM factors $\lambda_t^{(q)}$. 

\boldmath
\subsection{Effective Hamiltonian for $s\to d\ell^+\ell^-$}\label{sec:sdll}
\unboldmath

Let us recall that in the SM neglecting QCD corrections
 the top quark contribution to
 the effective Hamiltonian for $s\to d\ell^+\ell^-$ reads 
 \bea
\left[ \mathcal{H}_{\text{eff}}^{\ell\bar\ell}\right]^K_{\text{SM}}
&=& - g_{\text{SM}}^2 
\left[\lambda_t^{(K)} Y(x_t) \right] (\bar sd)_{V-A} 
(\bar\ell\ell)_{V-A}\nn\\
&&
+ 4 g_{\text{SM}}^2\sin^2\theta_W \left[\lambda_t^{(K)} Z(x_t) \right](\bar sd)_{V-A} 
(\bar\ell\ell)_{V}
 +h.c.
\label{eq:heffeins}
\eea
Here $Y(x_t)$ and $Z(x_t)$ are one-loop functions, analogous to $X(x_t)$, that
result from various penguin and box diagrams. The charm contributions and 
QCD corrections are irrelevant for the discussion presented below and will
be included only in the numerical analysis later on. We also remark 
that in principle also dipole operators could be included here, but {that} in
$K$ decays, as discussed in \cite{Buras:1994qa},  they can be fully neglected.
 Finally, the 
operator basis chosen in (\ref{eq:heffeins}) differs from the one used
to study QCD corrections \cite{Buras:1994qa}
 but is very suitable for the  discussion of
 modifications 
of the functions $Y(x_t)$ and $Z(x_t)$ due to NP contributions
 which we will discuss next.

Also in this case, $\left[ \mathcal{H}_{\text{eff}}^{\ell\bar\ell}\right]^K$
 receives tree level contributions of the gauge bosons $Z$, $Z^\prime$ and $Z_H$, and as now charged leptons appear in the final state, also the KK photon $A^{(1)}$ contributes.

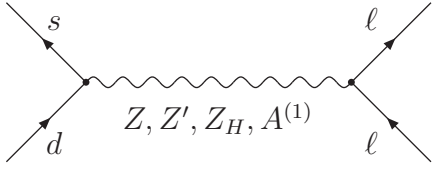
\begin{figure}
\begin{center}
\begin{picture}(150,100)(0,0)
\ArrowLine(10,10)(40,40)
\ArrowLine(40,40)(10,70)
\Photon(40,40)(140,40){2}{9}
\Vertex(40,40){1.3}
\Vertex(140,40){1.3}
\ArrowLine(140,40)(170,70)
\ArrowLine(170,10)(140,40)
\Text(90,20)[cb]{{\Black{$Z,Z^\prime,Z_H, A^{(1)}$}}}
\Text(28,60)[cb]{{\Black{$s$}}}
\Text(28,13)[cb]{{\Black{$d$}}}
\Text(148,60)[cb]{{\Black{$\ell$}}}
\Text(148,13)[cb]{{\Black{$\ell$}}}
\end{picture}
\end{center}
\caption{\it Tree level contributions of $Z,Z^\prime,Z_H$ and $A^{(1)}$ to the $s\to d\ell^+\ell^-$ effective Hamiltonian.}\label{Bleptons}
\end{figure}

 The relevant Feynman diagrams, shown in Fig.\ \ref{Bleptons},
contain on the l.\,h.\,s.\ the same vertices which we already encountered in the 
case of {the} $s\to d \nu \bar \nu$ decay. The relevant FCNC Lagrangian for
 $Z\bar sd$ couplings has been given in \eqref{eq:LeffZ1}.
For the $\ell^+ \ell^-$ vertex we write in analogy to (\ref{Lnunu})
\begin{equation}
\mathcal{L}_{\ell \bar \ell}(Z)= -\left[\Delta_L^{\ell\ell}(Z)(\bar\ell_L\gamma_\mu\ell_L)
+\Delta_R^{\ell\ell}(Z)(\bar\ell_R\gamma_\mu\ell_R)\right]
Z^\mu\,.
\end{equation}
The relevant entries have been collected in Appendix \ref{app:Deltas}.

The  evaluation of the $Z$-exchange in Fig.\ \ref{Bleptons} gives {then}
\bea
\left[ \Heff^{\ell\bar\ell}\right]_{Z}^K
&=& 
\frac{1}{M_Z^2}
 \left[ 
\Delta_L^{bs}(Z) (\bar s_L \gamma^{\mu} d_L)
+  \Delta_R^{bs}(Z) (\bar s_R \gamma^{\mu} d_R)
\right]  \nn\\
&&\hspace{1.5cm} \cdot \left[
 \Delta_L^{\ell\ell} (Z)(\bar \ell_L \gamma_{\mu} \ell_L)
+ \Delta_R^{\ell\ell} (Z)(\bar \ell_R \gamma_{\mu} \ell_R)\right]
 +h.c.\, ,
\label{eq:heffnew2}
\eea
which contains additional operators relative to \eqref{eq:heffeins}. The exchange of $Z^\prime$, $Z_H$ and $A^{(1)}$ gauge bosons yields analogous contributions that can simply be obtained from \eqref{eq:heffnew2} by replacing $Z$ by $Z^\prime$, $Z_H$ and $A^{(1)}$, respectively.

Following the previous discussion, we find that the effective Hamiltonian governing $s\to d\ell^+\ell^-$ transitions can be written in the compact form
\bea
\left[ \mathcal{H}_{\text{eff}}^{\ell\bar\ell}\right]^K
&=& - g_{\text{SM}}^2 
\left[ \lambda_t^{(K)} Y_K^{V-A} \right] (\bar sd)_{V-A} 
(\bar\ell\ell)_{V-A}\nn\\
&&
+ 4 g_{\text{SM}}^2\sin^2\theta_W \left[\lambda_t^{(K)} Z_K^{V-A} \right](\bar sd)_{V-A} 
(\bar\ell\ell)_{V}\nn\\
&&
- g_{\text{SM}}^2 
\left[\lambda_t^{(K)}  Y_K^V \right] (\bar sd)_{V} 
(\bar\ell\ell)_{V-A}\nn\\
&&
+ 4 g_{\text{SM}}^2\sin^2\theta_W \left[ \lambda_t^{(K)}  Z_K^V \right](\bar sd)_{V} 
(\bar\ell\ell)_{V}
 +h.c.\,,\label{eq:HeffKll}
\eea
where we have introduced the functions $Y_K^{V-A,V}$ and $Z_K^{V-A,V}$ defined as:
\bea\label{eq:YK}
Y_K^{V-A} &=& Y(x_t) +   \sum_{i=Z,Z^\prime,Z_H,A^{(1)}} (Y_{i}^K)^{V-A} \,,\\
Z_K^{V-A} &=& Z(x_t)+\sum_{i=Z,Z^\prime,Z_H,A^{(1)}} (Z_{i}^K)^{V-A} \,,\\
Y_K^{V} &=&   \sum_{i=Z,Z^\prime,Z_H,A^{(1)}} (Y_{i}^K)^{V} \,,\\
Z_K^{V} &=&  \sum_{i=Z,Z^\prime,Z_H,A^{(1)}} (Z_{i}^K)^{V}  \,,
\label{eq:tildeZK}
\eea
\noindent where 
\bea\label{eq:YKZ1}
(Y_{Z}^K)^{V-A} &=&
-\frac{1}{\lambda_t^{(K)}}\frac{\left[\Delta_L^{\ell\ell}(Z)-\Delta_R^{\ell\ell}(Z)\right]}{4M_{Z}^2g_\text{SM}^2}
\left[\Delta_L^{sd}(Z)-\Delta_R^{sd}(Z)\right]\,,\qquad\\
(Z_{Z}^K)^{V-A} &=&
\frac{1}{\lambda_t^{(K)}}\frac{\Delta_R^{\ell\ell}(Z)}{8M_{Z}^2g_\text{SM}^2\sin^2\theta_W}
\left[\Delta_L^{sd}(Z)-\Delta_R^{sd}(Z)\right]\,,\\
(Y_{Z}^K)^V &=& -\frac{1}{\lambda_t^{(K)}}\frac{\left[\Delta_L^{\ell\ell}(Z)-\Delta_R^{\ell\ell}(Z)\right]}{2M_{Z}^2g_\text{SM}^2} \Delta_R^{sd}(Z)\,,\\
(Z_{Z}^K)^V &=&
\frac{1}{\lambda_t^{(K)}}\frac{\Delta_R^{\ell\ell}(Z)}{4M_{Z}^2g_\text{SM}^2 \sin^2\theta_W} \Delta_R^{sd}(Z)\,.\label{eq:tildeZKZ1}
\eea
The $Z^\prime$, $Z_H$ and $A^{(1)}$ contributions can be straightforwardly obtained from \eqref{eq:YKZ1}--\eqref{eq:tildeZKZ1} by simply replacing $Z$ by $Z^\prime$, $Z_H$ and $A^{(1)}$, respectively.

\boldmath
\subsection{Effective Hamiltonian for $b\to d\ell^+\ell^-$ and  $b\to s\ell^+\ell^-$}\label{sec:bqll}
\unboldmath

Also in this case the effective Hamiltonian for $b\to q\ell^+\ell^-$ ($q=d,s$) can straightforwardly be obtained from \eqref{eq:HeffKll} by properly adjusting all flavour indices. In addition, in contrast to the $s\to d\ell^+\ell^-$ transition, now also the dipole operator {contributions} mediating the decay $b\to s\gamma$ become relevant.
The new RS contributions to the corresponding 
operators $\mathcal{Q}_{7\gamma}$ and
$\mathcal{Q}_{8G}$ appear first at the one-loop level and consequently as
already stated above are beyond the scope of this paper. Explicit formulae
for these contributions will be presented elsewhere. In the following we will
denote the total contribution of the dipole operators to the effective Hamiltonian in question simply by $\Heff(b\to s\gamma)$.

Adapting then the formula in \eqref{eq:HeffKll} to the present case, we find ($q=d,s$) 
\bea
\left[ \mathcal{H}_{\text{eff}}^{\ell\bar\ell}\right]^{B_q}
&=& \Heff(b\to s\gamma)- g_{\text{SM}}^2 
\left[\lambda_t^{(q)} Y_q^{V-A} \right] (\bar bq)_{V-A} 
(\bar\ell\ell)_{V-A} - g_{\text{SM}}^2 
\left[\lambda_t^{(q)}  Y_q^V \right] (\bar bq)_{V} 
(\bar\ell\ell)_{V-A}\nn\\
&&
+ 4 g_{\text{SM}}^2\sin^2\theta_W \left[\lambda_t^{(q)} Z_q^{V-A} \right](\bar bq)_{V-A} 
(\bar\ell\ell)_{V}
+ 4 g_{\text{SM}}^2\sin^2\theta_W \left[ \lambda_t^{(q)}  Z_q^V \right](\bar bq)_{V} 
(\bar\ell\ell)_{V}\nn\\
&&
+ h.c.\,.\label{eq:Heffqll}
\eea
In analogy to $Y_K^{V-A,V}, Z_K^{V-A,V}$ in
\eqref{eq:YK}--\eqref{eq:tildeZKZ1}, the relevant functions can be obtained
from the latter formulae by replacing $K$ by $q$. 
The same comment applies for the contributions of
$Z^\prime$, $Z_H$ and $A^{(1)}$.

\newsection{Exclusive Rare Decays}\label{sec:excl}

\boldmath
\subsection{$K^+ \rightarrow \pi^+\nu\bar\nu$ and $K_L \rightarrow \pi^0\nu\bar\nu$}
\unboldmath

Having at hand the effective Hamiltonian for $s\to d\nu\bar\nu$ transitions
derived in Section \ref{sec:sdnn} it is now straightforward to obtain
explicit expressions for the branching ratios $Br(\kpn)$ and $Br(\klpn)$.
 Reviews of these two decays can be found in 
\cite{Buras:2004uu,Isidori:2006yx,Smith:2006qg}.

As mentioned already in Section \ref{sec:sdnn}, now in addition to the usual SM {operator} $(\bar sd)_{V-A}(\bar\nu\nu)_{V-A}$ also the operator $(\bar sd)_{V}(\bar\nu\nu)_{V-A}$ is present. Therefore both matrix elements $\left<\pi^+|(\bar sd)_{V-A}|K^+\right>$ and $\left<\pi^+|(\bar sd)_{V}|K^+\right>$  have to be evaluated. Fortunately, as both $K^+$ and $\pi^+$ are pseudoscalar mesons, only the vector current part contributes and we simply have
\be
\left<\pi^+|(\bar sd)_{V-A}|K^+\right> = \left<\pi^+|(\bar sd)_{V}|K^+\right>\,.
\ee
This means that effectively, as in the LHT model, the effects of new
physics contributions can be collected in a single function that
generalises the SM one $X(x_t)$. Denoting this function as in \cite{Blanke:2006eb}
by
\be\label{eq:4.2}
X_K  \equiv X_K^{V-A} +X_K^V \equiv |X_K|e^{i\,\theta_X^K } \,,
\ee
we can make use of the formulae of Section 3.3 in \cite{Blanke:2006eb}  to
analyse the impact of new contributions on the branching ratios for $K^+
\rightarrow \pi^+\nu\bar\nu$ and $K_L \rightarrow \pi^0\nu\bar\nu$.
In particular we have
\begin{equation}\label{brknunu}
Br(\klpn)=r_1
(\sin\beta^K_X)^2 |X_K|^2,
\end{equation}
where 
\be\label{r1}
r_1=\kappa_L\left[\frac{\vts\vtd}{\lambda^5}\right]^2,\qquad
\beta^K_X=\beta-\beta_s -\theta_X^K\,,
\ee
with \cite{Mescia:2007kn }
\be
\kappa_L= (2.31\pm0.01)\cdot 10^{-10}\,,
\ee
and $\beta$ and $\beta_s$ defined through
\be\label{vtdvts}
V_{td}=\vtd e^{-i\beta}, \qquad V_{ts}=-\vts e^{-i\beta_s}\,.
\ee

Note that, in contrast to the real function $X(x_t)$, the new function 
$X_K$ is complex {implying} new CP-violating effects that can be best
tested in the very clean decay $K_L\to \pi^0\nu\bar\nu$.
 In this context the ratio 
\be
\frac{Br(K_{L}\to\pi^0\nu\bar\nu)}{Br(K_{L}\to\pi^0\nu\bar\nu)_\text{SM}}
=
\left|\frac{X_K}{X_\text{SM}}\right|^2\left[\frac{\sin\beta_X^K}{\sin{(\beta-\beta_s)}}\right]^2
\ee 
is very useful, as it is very sensitive to $\theta_X^K$ and is
theoretically very clean.

 The numerical analysis of both decays is presented in Section \ref{sec:num}.
In this context the most recent value $\kappa_+$ entering $Br(\kpn)$ is given
for $\lambda=0.226$ by \cite{Mescia:2007kn}
\be
\kappa_+=(5.36\pm0.026)\cdot 10^{-11}.
\ee
{The formulae for $Br(\kpn)$ can be found in \cite{Blanke:2006eb}.}

\boldmath
\subsection{$B \to K \nu \bar \nu$ and  $B \to K^* \nu \bar \nu$}
\unboldmath

{Since also the $B$ mesons are pseudoscalars,} following the arguments that led to (\ref{eq:4.2}) we easily find
\be\label{Ratio}
\frac{Br(B^+\to K^+\nu\bar\nu)}{Br(B^+\to K^+\nu\bar\nu)_{\rm SM}}=
\frac{|X_s|^2}{X(x_t)^2}\,,
\ee
where
\be
X_s\equiv X_s^{V-A}+X_s^V\equiv |X_s|e^{i\,\theta_X^s} \,.
\ee
The dilepton spectrum, sensitive only to $|X_s|$, can be found 
in equation (35) of \cite{BHI00}. Neglecting isospin breaking effects and 
$\Delta S=2$ CP-violating effects, one has
\be
Br(B^+\to K^+\nu\bar\nu)=2Br(B_d^0\to K_{L,S}\nu\bar\nu)\,.
\ee

The dilepton invariant mass spectrum of $B\to K^*\nu\bar\nu$ 
depends on two combinations of the relevant one loop functions so
that two ratios are of interest here:
\be
R_1=\frac{|X_s^{V-A}+X_s^V|^2}{X(x_t)^2}, \qquad 
R_2=\frac{|X_s^{V-A}|^2}{X(x_t)^2}.
\ee
The formula for the dilepton mass spectrum and the corresponding branching
ratio in terms of these two ratios can be found in equations (40)-(42) of
\cite{BHI00}. Unfortunately, the presence of three {form factors} introduces
some hadronic uncertainties. Therefore we will only present numerical results
for the ratio in (\ref{Ratio}) and $R_i$.

\boldmath
\subsection{$B_{d,s} \to \mu^+ \mu^-$}
\unboldmath

We will next consider the
 decays $B_{d,s} \to \mu^+ \mu^-$ that suffer from helicity suppression in the 
SM. This suppression cannot be removed through the exchanges of the gauge
 bosons in question but in principle could be removed through tree level 
exchanges of the Higgs boson. However the flavour conserving $H\mu\bar\mu$
vertex is proportional to the muon mass and in contrast to SUSY and general
two Higgs doublet models this suppression cannot be cancelled by a large
$\tan\beta$ enhancement. {In addition, flavour changing Higgs couplings
 receive a strong chirality 
suppression in addition to the usual RS-GIM suppression and are therefore 
negligibly small \cite{Blanke:2008zb}}.
In case of a bulk Higgs boson, also the Higgs KK
modes would contribute to $B_{d,s} \to \mu^+ \mu^-$, however also in this
latter case the $m_\mu$ suppression is effective. Therefore {in what follows} we restrict our
attention to the contributions of the SM $Z$ boson and 
heavy KK gauge bosons, calculated in Section \ref{sec:bqll}.

When evaluating the amplitudes for $B_{d,s} \to \mu^+ \mu^-$  by 
means of \eqref{eq:Heffqll} two simplifications occur. First when evaluating the matrix elements 
$\langle 0 | (\bar bq)_{V-A} | B_q\rangle$ and
$\langle 0 | (\bar bq)_{V} | B_q \rangle$ only the 
$\gamma_{\mu} \gamma_5$ part contributes as $B_q$ is pseudoscalar, so that
\be 
\langle 0 | (\bar bq)_{V} | B_q \rangle =0\,.
\ee
Then, due to the conserved vector current the vector component of the $\mu \bar \mu$-vertex drops out as well and 
 as in the SM
only the $\gamma_{\mu} \gamma_5$ component of the $\mu \bar \mu$-vertex is 
relevant. 
Therefore, since the dipole operator in $\mathcal{H}_{\text{eff}}(b\rightarrow s\gamma)$ does not contribute to this decay, the only operator contributing to $B_{d,s} \to \mu^+ \mu^-$ is the
SM $(V-A)\otimes (V-A)$ one, and the formulae of Section 3.4 of
\cite{Blanke:2006eb} can be applied here with {$Y_q$ replaced by $Y_q^{V-A}$ ($q=d,s$)},
where $Y_q^{V-A}$ {can be obtained from \eqref{eq:YK} by replacing ``$K$'' by ``$q$''}. In particular
\be
\frac{Br(B_q\to \mu^+\mu^-)}{Br(B_q\to \mu^+\mu^-)_{\rm SM}}=
\frac{|Y_q^{V-A}|^2}{Y(x_t)^2}.
\ee
This completes the analytic analysis of the $B_{s,d} \to \mu^+ \mu^-$ decays. 
The numerical results are discussed in Section~\ref{sec:num}.

\boldmath
\subsection{$K_L\to\mu^+\mu^-$}\label{sec:KLmumu}
\unboldmath
The discussion of the NP contributions to this decay is analogous
to $B_{d,s}\to \mu^+\mu^-$. Again only the SM operator $(V-A)\otimes (V-A)$
contributes and the real function $Y(x_t)$ is replaced by the complex
function {$Y_K^{V-A}\equiv |Y_K^{V-A}|e^{i\bar\theta_Y^K}$ defined in (\ref{eq:YK}). }

In contrast to the decays discussed until now, the short distance (SD)
contribution calculated here is only a part of a dispersive contribution
to $K_L\to\mu^+\mu^-$ that is by far dominated by the absorptive contribution
with two internal photon exchanges. Consequently the SD contribution
 constitutes only a small fraction of the 
branching ratio. Moreover, because of long distance contributions to the
dispersive part of $K_L\to\mu^+\mu^-$, the extraction of the short distance
part from the data is subject to considerable uncertainties. The most recent
estimate gives \cite{Isidori:2003ts}
\be\label{eq:KLmm-bound}
Br(K_L\to\mu^+\mu^-)_{\rm SD} \le 2.5 \cdot 10^{-9}\,,
\ee
to be compared with $(0.8\pm0.1)\cdot 10^{-9}$ in the SM 
\cite{Gorbahn:2006bm}.
In the model in question following \cite{Buras:2004aa} we
have ($\lambda=0.226$)
\be
Br(K_L\to\mu^+\mu^-)_{\rm SD} =
 2.08\cdot 10^{-9} \left[\bar P_c\left(Y_K\right)+
A^2 R_t\left|{Y^{V-A}_K}\right|\cos\bar\beta_{Y}^K\right]^2\,,
\ee
where we have defined:
\begin{gather}
\bar\beta_{Y}^K \equiv \beta-\beta_s-\bar\theta^K_Y\,,
\qquad \vtd=A\lambda^3 R_t\,,\\
\bar P_c\left(Y_K\right) \equiv \left(1-\frac{\lambda^2}{2}\right)P_c\left(Y_K\right)\,,
\end{gather}
with $P_c\left(Y_K\right)=0.113\pm 0.017$
\cite{Gorbahn:2006bm}. Here 
$\beta$ and $\beta_s$ are the phases of $V_{td}$ and $V_{ts}$ defined in 
(\ref{vtdvts}).
The numerical results are discussed in Section~\ref{sec:num}.

\boldmath
\subsection{$K_L\to\pi^0\ell^+\ell^-$}
\unboldmath
The rare decays $K_L\to\pi^0e^+e^-$ and $K_L\to\pi^0\mu^+\mu^-$ are dominated
by CP-violating contributions. The dominant indirect CP-violating
contributions are practically determined by the measured decays 
$K_S\to\pi^0 \ell^+\ell^-$ and the parameter $\varepsilon_K$. Consequently these
decays
are not as sensitive as $K_L\to \pi^0\nu\bar\nu$ to NP contributions
that are present here only in the subleading directly CP-violating contributions. 
Yet in models like the
LHT model with new sources of CP-violation enhancements of the branching
ratios by a factor of {1.5 can be found \cite{Blanke:2006eb,Blanke:2008ac}}.
 In this type of models, where
only the two SM operators in (\ref{eq:heffeins}) contribute, the effects 
of NP can be compactly summarised by generalisation of the 
real functions $Y(x_t)$ and $Z(x_t)$ to two complex functions $Y_K$ and 
$Z_K$, respectively.

In the model discussed here two new operators enter the game. Yet using
the same arguments as in the case of $K\to\pi\nu\bar\nu$ decays, we
 find that also here the two functions
\be
Y_K=Y_K^{V-A}+Y_K^V\,, \qquad  Z_K=Z_K^{V-A}+Z_K^V
\label{YKZK}
\ee
are sufficient to describe jointly the SM and NP contributions.
Consequently the formulae (8.1)--(8.8) of \cite{Blanke:2006eb} with 
$Y_K$ and $Z_K$ given in (\ref{YKZK}) can be used to study these decays
in the model in question. The original papers behind these formulae can 
be found in 
\cite{Buchalla:2003sj,Isidori:2004rb,Friot:2004yr,Mescia:2006jd,Buras:1994qa}.

Note that the presence of new operators is signalled by the additional
contributions $Y_K^V$ and $Z_K^V$ to $Y_K$ and $Z_K$, respectively.
Consequently, as no new operators enter the decay $K_L\to\mu^+\mu^-$, 
the functions $Y$ in the latter decay and in $K_L\to\pi^0\ell^+\ell^-$ 
differ from each other. This should be contrasted with the SM and the
LHT model, where they {are} {equal}.
The numerical results are discussed in Section~\ref{sec:num}.

\boldmath
\newsection{Inclusive Decays 
$B \to X_d \nu \bar \nu$ and  $B \to X_s \nu \bar \nu$}\label{sec:incl}
\unboldmath

Because of {the} right-handed couplings in the {$V q\bar s$ ($V=Z,Z_H,Z'$) vertices} the 
formulae (3.23)-(3.25) of \cite{Blanke:2006eb} for the inclusive decays $B \to X_{d,s}\nu\bar\nu$ have to be modified. There is no
interference between left-handed and right-handed contributions and
we find 
\be\label{brbnunu}
Br(B\to X_s\nu\bar\nu)=r_2\left(\left|X_s^{V-A}+\frac{X_s^V}{2}\right|^2+\left|\frac{X_s^V}{2}\right|^2\right),
\ee
where
\be
r_2=1.75~Br(B\to X_c e\bar\nu) 
\frac{3\alpha^2}{4\pi^2\sin^4\theta_W}
\frac{\vts^2}{\vcb^2} = (1.5\pm 0.2) \cdot 10^{-5},
\ee
with the factor $1.75$ summarising QCD and phase space corrections.

We find then
\be
\frac{Br(B\to X_s\nu\bar\nu)}{Br(B\to X_s\nu\bar\nu)_{\rm SM}}=
\frac{\left|X_s^{V-A}+\frac{X_s^V}{2}\right|^2+\left|\frac{X_s^V}{2}\right|^2}{X(x_t)^2}.
\ee
In the LHT model the second term in the numerator, that represents 
{the} $(V+A)$ contribution in the decomposition $(V-A)$ and $(V+A)$, is absent.

Of interest is also the ratio
\be\label{eq:P}
\frac{Br(B\to X_d\nu\bar\nu)}{Br(B\to X_s\nu\bar\nu)}=
\frac{\vtd^2}{\vts^2}\cdot P
\ee
where
\be
P\equiv
\frac{|X_d^{V-A}+X_d^V/2|^2+|X_d^V/2|^2}
{|X_s^{V-A}+X_s^V/2|^2+|X_s^V/2|^2}.
\ee
In the SM and models with {Constrained Minimal Flavour Violation (CMFV) \cite{Buras:2000dm,Buras:2003jf,Blanke:2006ig}, in which all flavour violation is governed by the CKM matrix and only SM operators are relevant\footnote{See \cite{D'Ambrosio:2002ex,Chivukula:1987py,Hall:1990ac} for a more general definition of Minimal Flavour Violation (MFV), in which new operators are allowed.},} one has $P=1.$ Note that (\ref{eq:P}) with
$P=1$
represents one of many correlations in models with CMFV to which we will
now turn our attention.

In the SM and in models with CMFV  there is {also} a striking correlation between the 
branching ratios
for $\klpn$ and $B\to X_s\nu\bar\nu$ as the same one-loop function
$X(x_t)$ governs the two processes in question \cite{Buras:2001af}. 
This relation is generally
modified in models with non-CMFV interactions. As this modification  beyond 
CMFV has not
been discussed in the literature we will present it here.
Using (\ref{brknunu}) and (\ref{brbnunu}) we find
\be\label{eq:rel-nn}
\frac{Br(\klpn)}{Br(B\to X_s\nu\bar\nu)}=\frac{r_1}{r_2}(\sin\beta^K_X)^2
\frac{|X_K|^2}{\left|X_s^{V-A}+\frac{X_s^V}{2}\right|^2+\left|\frac{X_s^V}{2}\right|^2}\,,
\ee
{which reduces in CMFV models to}
\be
\frac{Br(\klpn)}{Br(B\to X_s\nu\bar\nu)}=\frac{r_1}{r_2}\sin(\beta-\beta_s)^2\,.
\ee

\newsection{Correlations Between Various Observables}\label{sec:correlations}

In the SM and in models with CMFV the rare decays analysed in the present 
paper depend basically on three universal functions $X$, $Y$, $Z$. 
Consequently, a number of correlations exist between various observables
not only within the $K$ and $B$ systems but also between $K$ 
and $B$ systems. In particular the latter correlations are very interesting
as they are characteristic for this class of models. A review of these
correlations is given in \cite{Buras:2003jf}. As already stressed several 
times in our paper these correlations are violated in the model considered.
Such violations have also been found in the LHT model \cite{Blanke:2006eb}.

In our numerical analysis in Section \ref{sec:num} we will investigate 
a multitude of correlations, giving there relevant formulae if necessary. {One has {already} been given in \eqref{eq:rel-nn}.}
One can distinguish the following classes of correlations:

{\bf Class 1:}
 Correlations implied by the universality of the real function $X$ {in CMFV models}. They
 involve rare $K$ and $B$ decays with $\nu\bar\nu$ in the final state.

 {\bf Class 2:} 
 Correlations implied by the universality of the real function $Y$ {in CMFV models}. They
 involve rare $K$ and $B$ decays with $\mu^+\mu^-$ in the final state.

{\bf Class 3:} 
 In models with CMFV NP contributions enter the functions $X$ and $Y$
 approximately in the same manner as at least in the Feynman gauge
 they come dominantly from Z penguin diagrams. This implies 
 correlations between
 rare decays with $\mu^+\mu^-$ and $\nu\bar\nu$ in the final state.
 It should be emphasised that this is a separate class as NP can generally
 have {a} different impact on decays with $\nu\bar\nu$ {and} $\mu^+\mu^-$ in the
 final state.

 {\bf Class 4:}
 Here we group correlations between $\Delta F=2$ and $\Delta F=1$ transitions
 in which the one-loop functions $S$ and $(X,Y)$, respectively, cancel out and
 the correlations follow from the universality of the CKM parameters. The two best
 known correlations of this type are two {\it golden} relations 
\cite{Buchalla:1994tr,Buras:2001af,Buras:2003td} that we will analyse in 
Section \ref{sec:golden}.

 {\bf Class 5:}
 Here we group correlations within $\Delta F=2$ transitions. The best known
 is the one between the asymmetries $S_{\psi\phi}$ and $A^s_\text{SL}$
 \cite{Ligeti:2006pm} analysed by us already in \cite{Blanke:2008zb}.

 As we will see in Section \ref{sec:num}, some of these correlations, in
 particular those between $K$ and $B$ decays are strongly violated, others
 are approximately satisfied. Clearly the full picture is only obtained by looking
 simultaneously at patterns of violations of the correlations in question in a
given NP scenario.

{
\begin{table}[ht]
\renewcommand{\arraystretch}{1}\setlength{\arraycolsep}{1pt}
\center{\begin{tabular}{|l|ccc|}
\hline
Class & \multicolumn{3}{|c|}{Correlated decays/observables}\\
\hline
 1 & $K_L\to \pi^0 \nu\bar\nu$ & $\longleftrightarrow$&  $K^+\to \pi^+ \nu\bar\nu$\\
 & $K_L\to \pi^0 \nu\bar\nu$ & $\longleftrightarrow$&$B\to X_{s,d}\nu\bar\nu$\\
  & $B\to X_s\nu\bar\nu$      & $\longleftrightarrow$&$B\to X_d\nu\bar\nu$\\
\hline
2  & $K_L\to \pi^0 \mu^+\mu^-$ & $\longleftrightarrow$&$K_L\to \pi^0 e^+e^-$\\
 & $K_L\to\mu^+\mu^-$        & $\longleftrightarrow$&$B_s\to\mu^+\mu^-$\\
  &  $B_s\to\mu^+\mu^-$        & $\longleftrightarrow$&$B_d\to\mu^+\mu^-$\\
\hline
 3 & $K_L\to \pi^0 \nu\bar\nu$ & $\longleftrightarrow$&$K_L\to \pi^0 \mu^+\mu^- \left(e^+e^-\right)$\\
 & $\kpn$ & $\longleftrightarrow$& $K_L\to\mu^+\mu^-$\\
& $K_L\to \pi^0 \nu\bar\nu$ & $\longleftrightarrow$&$B_s\to\mu^+\mu^-$\\
  & $B\to X_s\nu\bar\nu$     & $\longleftrightarrow$&$B_s\to\mu^+\mu^-$\\
\hline
4 & $B_{s,d}\to\mu^+\mu^-$     & $\longleftrightarrow$&$\Delta M_{s,d}$\\
  & $K\to \pi\nu\bar\nu$      & $\longleftrightarrow$&$S_{\psi K_S}$\\
\hline
5 & $S_{\psi\phi}$            & $\longleftrightarrow$&$A^s_\text{SL}$\\
\hline 
\end{tabular}  }
\caption {\textit{Examples of the several classes of correlations.}}
\label{tab:correlations}
\renewcommand{\arraystretch}{1.0}
\end{table}
}

 In Table \ref{tab:correlations} we collect examples of correlations in each class that constitute the
 most powerful tests of NP. Needles to say the classification of correlations presented here
 is valid for any extension of the SM.

\boldmath\newsection{Anatomy of Contributions of $Z$, $Z_H$ and $Z^\prime$ Gauge
  Bosons}\label{sec:anatomy}
\unboldmath
The discussion of the last four sections was rather general and the
formulae given there can easily be adapted to any model with tree level
heavy neutral gauge boson exchanges. We will now turn to the specific model
considered here, beginning with an anatomy of various contributions.

The  NP contributions to {the} functions $X$, $Y$ and $Z$ given in the
previous section are a product of three main components: the coupling of the
respective gauge boson to the down-type quarks, its propagator in the low
energy limit, and finally the gauge boson's coupling to leptons. 
 For a given meson system characterised by $(ij)$ there are six distinct contributions
from the three gauge bosons $Z$, $Z_H$ and $Z^\prime$ coupling to left- and right-handed down-type quarks, $\Delta_{L,R}^{ij}(Z)\,,\Delta_{L,R}^{ij}(Z_H)\,,\Delta_{L,R}^{ij}(Z^\prime)$. Two of them, the couplings of $Z$ and $Z^\prime$ to the left-handed quarks are suppressed by the custodial {symmetry}. To understand the relative sizes of these six contributions, it is necessary to investigate the hierarchies in the above mentioned building blocks as we will do in the following.

{We note that in case of the $Y$ and $Z$ functions also the KK photon $A^{(1)}$ contributes. However its couplings to fermions are suppressed by the smallness of the electromagnetic coupling $e^\text{4D}$ and the electric quark charge, so that its contributions turn out to be small {(if not absent)} in all cases.}

\subsection{Couplings to Quarks}

For the gauge couplings to left-handed quarks the hierarchy is given by the mixing of gauge bosons
into mass eigenstates (see \eqref{A.1}, \eqref{A.2}, \eqref{A.16}) and by the suppression induced by the custodial protection. Numerically, we find
\begin{equation}
\Delta_L^{ij}(Z_H):\Delta_L^{ij}(Z^\prime):\Delta_L^{ij}(Z)\sim\mathcal{O}(10^4):{\mathcal{O}(10^3)}:1 \,.
\end{equation}
For the couplings to the right-handed quarks, the hierarchy is solely determined by the mixing of gauge bosons into mass eigenstates, and is given by
\begin{equation}
\Delta_R^{ij}(Z_H):\Delta_R^{ij}(Z^\prime):\Delta_R^{ij}(Z)\sim\mathcal{O}(10^2):\mathcal{O}(10^2):1 \,,
\end{equation}
where these hierarchies hold for the $K$, $B_d$ and $B_s$ systems likewise,
that is for {$ij=sd$, $ij=bd$ and $ij=bs$}, respectively.

{We note that in the presence of an exact protective $P_{LR}$ symmetry the flavour violating couplings $\Delta_L^{ij}(Z)$ and $\Delta_L^{ij}(Z')$ would vanish identically.
In this limit the same linear combination of $Z^{(1)}$ and $Z_X^{(1)}$ enters the $Z$ and $Z'$ mass eigenstates, so that the same cancellation of contributions is effective. Taking 
into account the $P_{LR}$-symmetry breaking effects on the UV brane, the custodial protection mechanism is not exact anymore, but still powerful enough to suppress $\Delta_L^{ij}(Z)$ by two orders of magnitude.
In the case of $Z'$, the mixing angles for $Z^{(1)}$ and $Z_X^{(1)}$ are modified by roughly 10\% when including the violation of the $P_{LR}$ symmetry \cite{Blanke:2008aa}. Accordingly, the protection is weaker in the case of $Z'$ and $\Delta_L^{ij}(Z')$ is suppressed only by one order of magnitude 
compared to the case without protection.

As the right-handed down-type quarks are no $P_{LR}$-eigenstates, the custodial protection mechanism is not effective in the case of $\Delta_R^{ij}(Z)$ and $\Delta_R^{ij}(Z')$, which explains the different pattern of hierarchies in the right-handed sector.
}

{This general picture is unaffected by the inclusion of the effects of KK fermion mixing.}

\subsection{Gauge Boson Propagators}
If we assume the additional neutral gauge bosons $Z_H$ and $Z^\prime$ to be degenerate in mass, as done in (\ref{degeneracymasses}), their contribution to the functions $X$, $Y$ and $Z$ is suppressed by a factor $M_{Z}^2/M_{\text{KK}}^2\sim\mathcal{O}(10^{-3})$ with respect to the $Z$ contribution.

\subsection{Couplings to Leptons}
For this comparison, we assume the lepton zero mode localisation to be flavour independent, that is we assume degenerate bulk masses in the lepton sector. Since leptons are significantly lighter than quarks of the same generation, we choose them to be localised towards the UV brane and set the bulk mass parameters to $c=\pm 0.7$ for left- and right-handed leptons, respectively. {This assumption is well motivated by the observation that the flavour conserving couplings depend only very weakly on the actual value of $c$, provided $c>0.5$ ($c<-0.5$ for right-handed fermions).} Since the couplings of gauge boson mass eigenstates are dominated by the $Z^{(0)}$ and $Z^{(1)}$ contributions\footnote{This is due to the fact that the overlap integral of a $(++)$ gauge boson with UV localised fermions is much larger than the corresponding overlap integral for a $(-+)$ gauge boson.}, their hierarchy does not depend on the particular handedness or species of leptons involved. In contrast to the $Z_H$ and $Z^\prime$ coupling, the $Z$ coupling to the lepton sector is not suppressed by an overlap integral of shape functions and hence is expected to be dominant. Numerically, 
\begin{equation}
\Delta_{L,R}^{\nu\nu,\ell\ell}(Z_H):\Delta_{L,R}^{\nu\nu,\ell\ell}(Z^\prime):\Delta_{L,R}^{\nu\nu,\ell\ell}(Z)\sim\mathcal{O}(10^{-1}):\mathcal{O}(10^{-1}):1\,.
\end{equation}
This hierarchy is obviously the same in $K$, $B_d$ and $B_s$ systems.

\subsection{Putting Together the Building Blocks}
The above considerations now can be used to weight the contributions of $Z$, $Z_H$ and $Z^\prime$ coupling to left- and right-handed quarks. It is obvious that the contributions from the $Z_H$ and $Z$ coupling to left-handed quarks are comparable in size, while the corresponding contribution from $Z^\prime$ is clearly negligible. The contribution from couplings to right-handed quarks is strictly dominated by the $Z$ gauge boson. To finally determine the dominant overall contribution, we note that due to the custodial protection and the particular structure of the model the $Z$ boson couples much more strongly to right-handed quarks than to left-handed quarks, $\Delta_R^{ij}(Z)\gg\Delta_L^{ij}(Z)$, which is even more the case if we concentrate on parameter sets that can produce significant modifications to the functions $X$, $Y$ and $Z$.

The main message from our semi-analytic analysis is the following: If the
effects in rare $K$ and $B$ decays are significant, they are dominantly caused
by the $Z$ boson coupling to {\it right-handed} {down} quarks.

\boldmath
\subsection{Comparison of $K$ and $B_{d,s}$ Systems}
\unboldmath

\begin{figure}
\begin{minipage}{7.9cm}
{\psfig{file=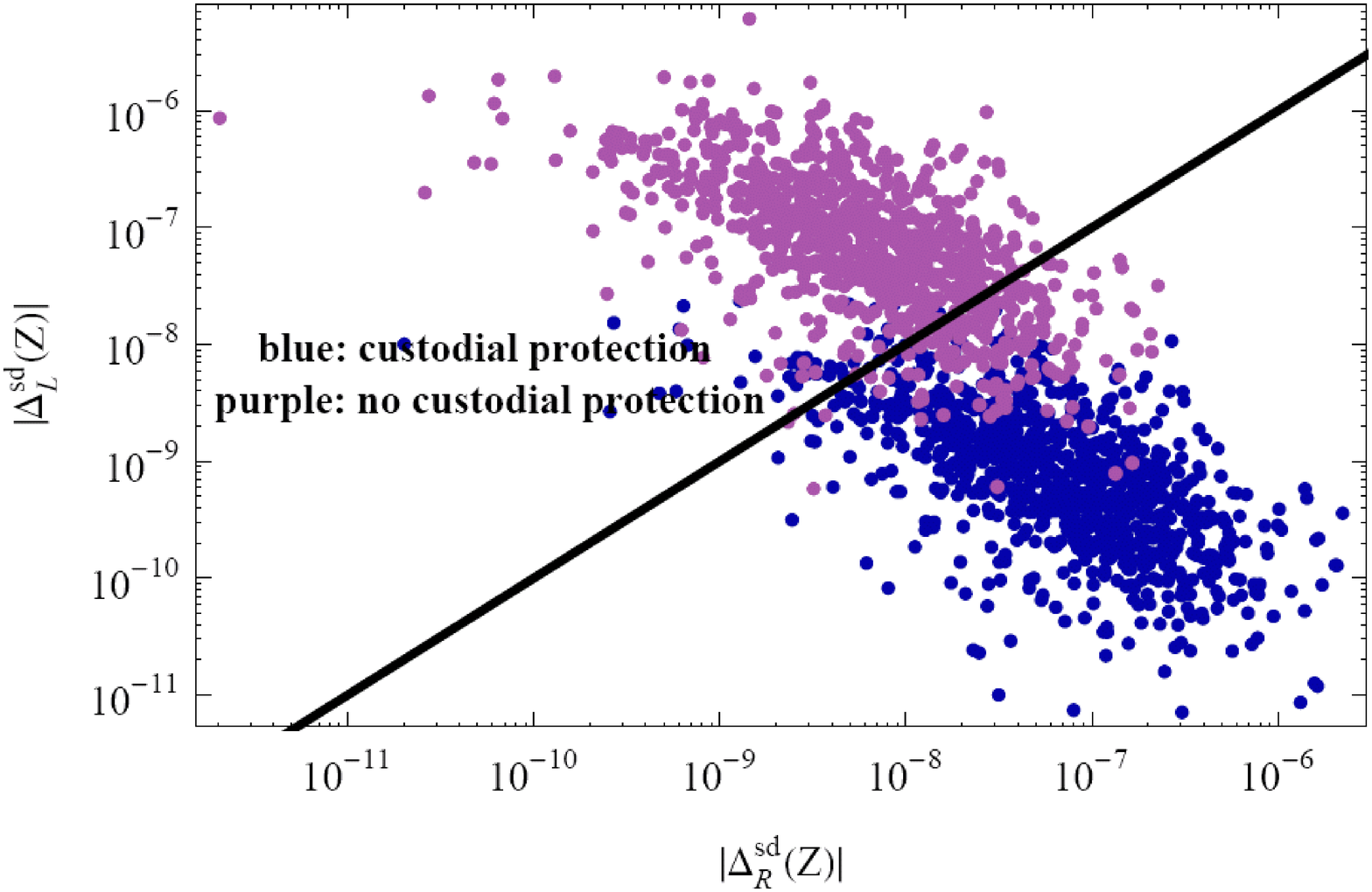,scale=.22}}
\end{minipage}
\begin{minipage}{7.7cm}
\center{\psfig{file=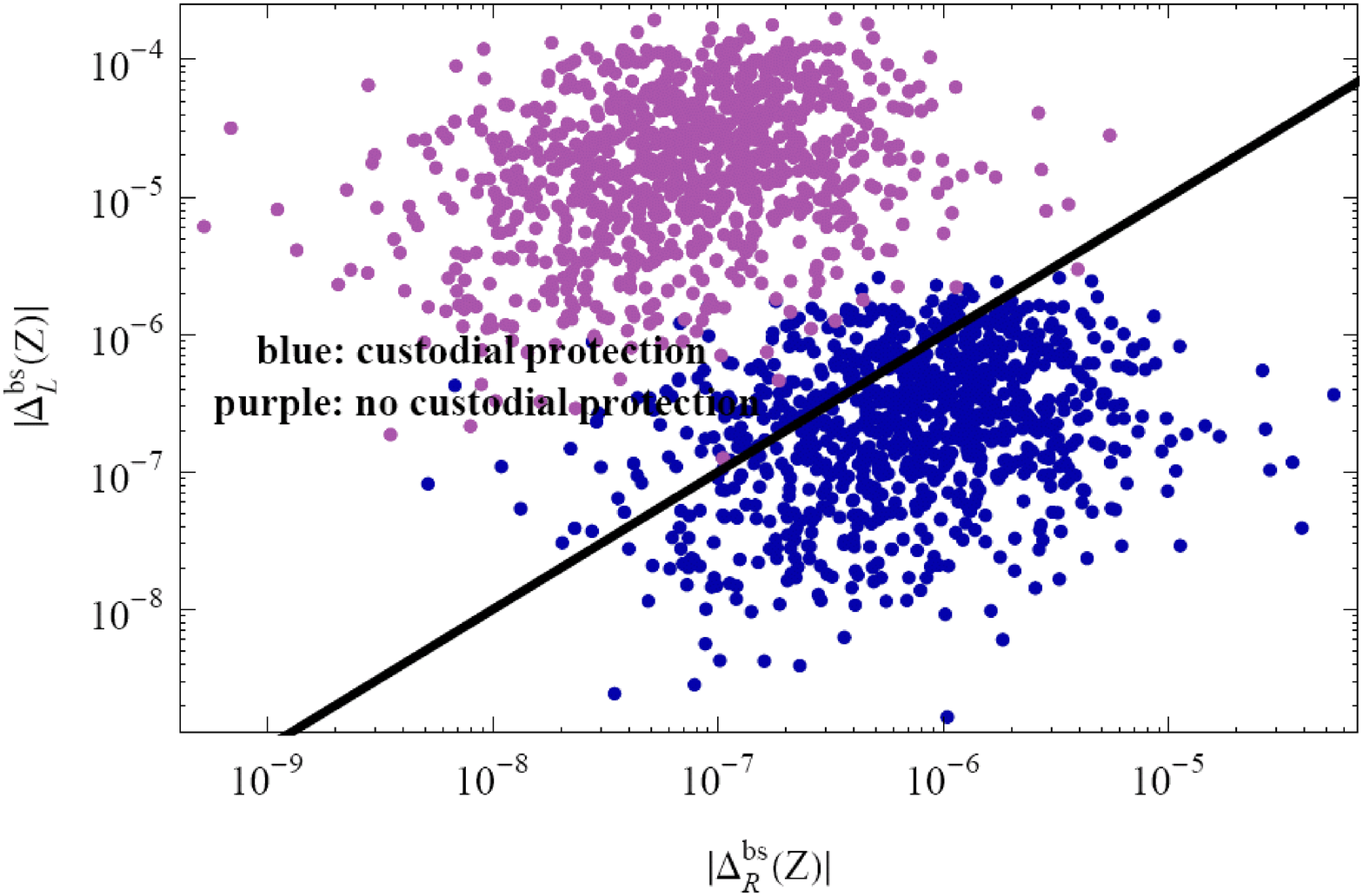,scale=.22}}
\end{minipage}
\caption{\it $|\Delta_L^{ij}(Z)|$ versus $|\Delta_R^{ij}(Z)|$ for $ij=sd$ (left) and $ij=bs$ (right). The blue points are obtained in the custodially protected model after imposing all constraints from $\Delta F=2$ observables \cite{Blanke:2008zb}. The purple points show the effect of removing the custodial protection, as outlined in Section \ref{sec:remove}. The solid lines display the equality  $|\Delta_L^{ij}(Z)| = |\Delta_R^{ij}(Z)|$.
\label{fig:Deltas}}
\end{figure}

As the tree level $Z$ contributions turn out to be dominant, {from now on we restrict our discussion} to these contributions. In Fig.\ \ref{fig:Deltas} we show the
allowed ranges for $\Delta_{L,R}^{sd}(Z)$ and $\Delta_{L,R}^{bs}(Z)$,
respectively. The solid thick line corresponds to the equality of left- and
right-handed couplings.
We observe:
\begin{itemize}
\item
$\Delta_{R}^{sd}(Z)$ is larger than $\Delta_{L}^{sd}(Z)$ for a dominant part
  of the allowed points and is {on  average} larger than $\Delta_{L}^{sd}(Z)$
  by two orders of magnitude.
\item
The dominance of $\Delta_{R}^{bs}(Z)$ over $\Delta_{L}^{bs}(Z)$ is less
pronounced, but still {on average} $\Delta_{R}^{bs}(Z)$ is larger than
$\Delta_{L}^{bs}(Z)$ by one order of magnitude.
\item
The values of $\Delta_{R}^{bs}(Z)$ are {on average} larger than
$\Delta_{R}^{sd}(Z)$ by one order of magnitude{, as the $(b_R,s_R)$ system is localised closer to the IR brane than the $(s_R,d_R)$ system.}
\end{itemize}
For $\Delta_{L,R}^{bd}(Z)$ we find the values between those for the $(bs)$ and
$(sd)$ cases.

When comparing the size of NP effects in $K$ and $B_{d,s}$ systems {we have to} take into account that 
 NP
contributions are also enhanced non-universally by  factors $1/\lambda_t^{(i)}$.
As  $\lambda_t^{(K)}\simeq 4\cdot 10^{-4}$, whereas
$\lambda_t^{(d)}\simeq 1\cdot 10^{-2}$ and $\lambda_t^{(s)}\simeq
4\cdot 10^{-2}$, we would naively expect the deviation from the SM functions
in the
$K$ system to be by an order of magnitude larger than in the
$B_d$ system, and even by a larger factor than in the
$B_s$ system.  This strong hierarchy in the factors $1/\lambda_t^{(i)}$ is partially compensated by the opposite hierarchy in $\Delta_R^{ij}(Z)$. However, as flavour violation is generally weaker in the right-handed sector, this compensation is not complete, so that still larger effects are expected in $K$ physics than in $B_{d,s}$ physics. In any case the universality for the functions $X$, $Y$ and $Z$ in the $K$ and $B$ systems is necessarily broken.

 Having at hand numerical results for $\Delta_R^{sd,bd,bs}(Z)$ for a large
 number of parameter sets, 
we can  predict the average relative size of NP contributions in the $K$ and
 $B$ systems. 
We find that the size of the NP contributions on average drops by a factor of
 four when going from the 
$K$ to the $B_d$ system and by another factor of two when going from the $B_d$
 to the $B_s$ system.

\boldmath
\subsection{Removing the Protection of Left-Handed $Z$ Couplings}\label{sec:remove}
\unboldmath

It is instructive to investigate how our results would look like if
the protection of the left-handed $Z$ couplings to {down-type} quarks was
not present. In order to get a rough idea we simply removed the 
contributions of the $Z^{(1)}_X$ gauge boson to the {$Z$, $Z_H$ and $Z'$} couplings that
are generated in the process of electroweak symmetry breaking.
{This also has an impact on the right-handed $Z$ couplings, as those were dominated by the $Z_X^{(1)}$ contribution and are now suppressed by a factor $\sin^2\theta_W\simeq 0.2$. However} the main
effect is the enhancement of the couplings $\Delta_L^{ij}(Z)$ by
roughly two orders of magnitude. {The results are displayed by the purple points in Fig.\ \ref{fig:Deltas} 
for the $K$ and $B_s$ systems}. We observe that
$\Delta_L^{sd}(Z)$ tends now to be larger than $\Delta_R^{sd}(Z)$, 
while $\Delta_L^{bs}(Z)$ fully dominates over $\Delta_R^{bs}(Z)$.
Again intermediate results are found for $\Delta_{L,R}^{bd}(Z)$.

{It is important to note that now, as the rare decays in question are fully dominated by the $\Delta_L^{ij}(Z)$ contribution, the expected pattern of deviations from the SM changes drastically with respect to the custodially protected scenario. As $\Delta_L^{ij}(Z)$ exhibit a similar hierarchy as the CKM factors $\lambda_t^{(q)}$, relative NP effects of roughly equal size are expected in $K$ and $B$ decays. We stress however that a more quantitative analysis in that case requires also the inclusion of the $Z b_L\bar b_L$ constraint, possibly altering the pattern of expected effects. 
{In addition, removing the $Z_X^{(1)}$ couplings also modifies the predictions for $\Delta B=2$ observables at the $\ord(100\%)$ level, so that the points from our parameter scan do in general not fulfil the associated constraints any more. On the other hand the most severe constraint comes from $\eps_K$, which we have shown in \cite{Blanke:2008zb} to be dominated by KK gluon contributions and thus insensitive to the precise structure of the EW sector. Consequently we do not expect our results to be affected significantly by this simplified working assumption.
}

In the next section we will show a couple of examples {of how removing the protection in question influences} rare decay branching ratios.

\newsection{Numerical Analysis}\label{sec:num}
\label{sec:numerics}
\subsection{Preliminaries}
\label{subsec:5.1}

In our numerical analysis we will set $|V_{us}|$, $|V_{cb}|$ and
$|V_{ub}|$ to their central values measured in tree level
decays and collected in
Table~\ref{tab:input}.

\begin{table}[ht]
\renewcommand{\arraystretch}{1}\setlength{\arraycolsep}{1pt}
\center{\begin{tabular}{|l|l|}
\hline
$\lambda=|V_{us}|= 0.226(2)$ & $G_F= 1.16637\cdot 10^{-5}\gev^{-2}$ \qquad {} \\
$|V_{ub}| = 3.8(4)\cdot 10^{-3}$ &  $M_W = 80.403(29) \gev$ \\
$|V_{cb}|= 4.1(1)\cdot 10^{-2}$ \hfill\cite{Bona:2006sa}& $\alpha(M_Z) = 1/127.9$ \\\cline{1-1}
$ \gamma = 75(25)^\circ $  & $\sin^2\theta_W = 0.23122$\\\cline{1-1}
$\Delta M_K= 0.5292(9)\cdot 10^{-2} \,\text{ps}^{-1}$ \qquad {} & $m_K^0= 497.648\mev$ \\
$|\eps_K|= 2.232(7)\cdot 10^{-3}$ \hfill\cite{Amsler:2008zz}& $m_{B_d}= 5279.5\mev$ \\\cline{1-1}
$\Delta M_d = 0.507(5) \,\text{ps}^{-1}$ & $m_{B_s} = 5366.4\mev$ \hfill\cite{Amsler:2008zz} \\\cline{2-2}
$\Delta M_s = 17.77(12) \,\text{ps}^{-1}$  & $\eta_1= 1.32(32)$ \hfill\cite{Herrlich:1993yv}\\\cline{2-2}
$S_{\psi K_S}= 0.671(24)$ \hfill\cite{Barberio:2007cr}&  $\eta_3=0.47(5)$ \hfill\cite{Herrlich:1995hh,Herrlich:1996vf} \\\hline
$\bar m_c = 1.30(5)\gev$ & $\eta_2=0.57(1)$ \\
$\bar m_t = 162.7(13)\gev$ & $\eta_B=0.55(1)$ \hfill \cite{Buras:1990fn,Urban:1997gw} \\\hline
$F_K = 156(1)\mev$ \hfill \cite{Flavianet}& $F_{B_s} = 245(25)\mev$ \\\cline{1-1}
$\hat B_K= 0.75(7)$ & $F_{B_d} = 200(20)\mev$ \\
$\hat B_{B_s} = 1.22(12)$ & $F_{B_s} \sqrt{\hat B_{B_s}} = 270(30)\mev$ \\
$\hat B_{B_d} = 1.22(12)$ & $F_{B_d} \sqrt{\hat B_{B_d}} = 225(25)\mev$ \\
$\hat B_{B_s}/\hat B_{B_d} = 1.00(3)$ \hfill \cite{Lubicz:2008am}& $\xi = 1.21(4)$ \hfill \cite{Lubicz:2008am} \\\hline
$\tau(B_s)=1.470(26) \,\text{ps}$ & {$\alpha_s(M_Z)=0.118(2)$} \hfill\cite{Amsler:2008zz}\\
 $\tau(B_d)= 1.530(9) \,\text{ps}$\hfill\cite{Amsler:2008zz}&\\\hline
\end{tabular}  }
\caption {\textit{Values of the experimental and theoretical
    quantities used as input parameters.}}
\label{tab:input}
\renewcommand{\arraystretch}{1.0}
\end{table}

As the fourth parameter we choose the angle $\gamma$ of the
standard unitarity triangle that to an excellent approximation equals the
phase $\delta_\text{CKM}$ in the CKM matrix. The angle $\gamma$ has been 
extracted
from $B \to D^{(*)} K$ decays without the influence of 
NP. {The value used throughout our analysis and quoted in Table~\ref{tab:input} is consistent with recent fit results \cite{Bona:2006sa,Charles:2004jd}.}

The {``true''} value of $\beta$ is obtained from
\be
 R_b=\left(1-\frac{\lambda^2}{2}\right)\frac{1}{\lambda}\frac{|V_{ub}|}{\vcb}= 0.40 \pm 0.04
\ee
and  $\gamma$, i.\,e. from tree level decays only and is
not affected by {a potential} NP phase. We find then
\begin{equation}
 (\sin 2\beta)_{\rm true}= (0.726 \pm 0.070),\qquad \beta_{\rm true}=(23.3 \pm 2.9)^\circ\,,
\end{equation}
that is consistent with  $S_{\psi K_S}$ in Table~\ref{tab:input}, although
a bit larger
implying a  small negative value of a NP phase {$\varphi_{B_d}$
 in  $B_d-\bar B_d$} mixing:
\be
S_{\psi K_S}=\sin(2\beta_{\rm true}+2{\varphi_{B_d}}),\qquad {\varphi_{B_d}}= - (2.2 \pm 3.1)^\circ\,,
\ee
 as discussed already by several authors in the literature. This new phase
 can be easily obtained in the model considered \cite{Blanke:2008zb}.

As pointed out recently in \cite{Buras:2008nn}, the value of $S_{\psi K_S}$ in 
Table~\ref{tab:input} and even the value of $(\sin 2\beta)_{\rm true}$
given above appear too small to obtain the experimental value of the 
CP-violating parameter $\varepsilon_K$ in the SM. Similar tensions
between CP-violation in $K^0-\bar K^0$ and $B_d^0-\bar B_d^0$ mixings 
from a different point of view have been pointed out in \cite{Lunghi:2008aa}. 
All these
tensions can be removed in the model considered.

For the non-perturbative parameters entering the analysis of
particle-antiparticle mixing we choose and collect in Table~\ref{tab:input}
their lattice averages given in \cite{Lubicz:2008am}.

In order to simplify our numerical analysis we will, as in \cite{Blanke:2008zb}, 
set all non-perturbative 
parameters to their central values and instead we will allow $\Delta M_K$, 
$\varepsilon_K$, $\Delta M_d$, $\Delta M_s$ and $S_{\psi K_S}$ to differ from 
their experimental values by $\pm 50\%$, $\pm 20\%$, $\pm 30\%$, $\pm 30\%$ 
and $\pm 20\%$, respectively. In the case of $\Delta M_s/\Delta M_d$ we
will choose $\pm 20\%$, as the error on the relevant parameter, $\xi$, is
smaller than in the case of $\Delta M_d$ and $\Delta M_s$ separately.
 The relevant expressions for these observables  within the
model considered {are given} in
\cite{Blanke:2008zb}. These uncertainties could appear rather conservative, but we 
do not want to miss any interesting effect by choosing too optimistic 
non-perturbative uncertainties.

 In presenting the results below we impose all existing constraints
 from $\Delta F=2$ transitions analysed by us in \cite{Blanke:2008zb} and 
 require that  all quark masses and weak mixing angles calculated
 in this model agree with the experimental ones
 within $2\sigma$. The details behind this latter calculation are
 given in \cite{Blanke:2008zb}. {Specifically we use the parameterisation of the 5D Yukawa couplings presented in that paper, where we scan over $0\le y^i_{u,d}\le 3$ in order to maintain perturbativity, and vary the relevant mixing angles and CP-violating phases in their physical ranges $[0,\pi/2]$ and $[0,2\pi]$, respectively. For the bulk mass parameters we impose $0.1\le c_Q^3\le0.5$ and fit the other values in order to obtain correct quark masses and CKM mixing angles. For further details on the parameter scan we refer the reader to \cite{Blanke:2008zb}.}

As there is some fine-tuning required to 
 fit the experimental value of $\varepsilon_K$ we will consider as our main
 results for rare decays those obtained from
  points in the parameter space for which this fine-tuning is
 moderate and characterised by the Barbieri-Giudice {\cite{Barbieri:1987fn}} measure 
$\Delta_{\rm BG}(\varepsilon_K)\le 20$. 
They are given by {\it orange} points in the plots below.
However, for completeness we will also
 show results obtained for {arbitrarily} high fine-tuning. They are represented
by {\it blue} points in the figures below and obviously show {on average} larger
deviations from the SM than the ones found with only moderate {fine-tuning}. To be specific, all the statements
from now on apply only to the latter points. {For some examples we also show the results obtained after removing the custodial protection. In that case points with {arbitrarily} high fine-tuning are shown in {\it purple}, while points with moderate fine-tuning are {shown} in {\it green}.}

We will {perform} the numerical analysis in the 
same spirit as in the LHT model so that an easy comparison of
the results obtained in the LHT model {\cite{Blanke:2006eb,Blanke:2008ac}} and the results in the  model
 discussed here will be possible. Therefore the presentation below follows
closely subsections 10.4--10.11 of \cite{Blanke:2006eb}, {although it contains new correlations that cannot be found in \cite{Blanke:2006eb}.}

\subsection{Breakdown of Universality}
\label{subsec:UB}

In {CMFV} models the functions $X_i$, $Y_i$ and $Z_i$ are {real and} independent of the
index  $i=K,d,s$. Consequently, they are universal quantities implying strong 
correlations between observables in $K$, $B_d$ and $B_s$ systems. In the
model discussed here this universality is generally broken, as clearly seen in {the}
formulae of Sections \ref{sec:excl} and \ref{sec:incl}. Moreover the
 functions $X_i$, $Y_i$ and $Z_i$ become complex quantities and their phases
turn out to exhibit {a} non-universal behaviour.

{To get a feeling for the possible sizes of $|X_i|$, $|Y_i|$ and $|Z_i|$ $(i=K,d,s)$ we give the 5$\sigma$ ranges for the distribution of the respective quantity. To also capture non-symmetric distributions around the mean value, for each quantity we determine the standard deviations for two symmetrised versions of the distribution: one that originates from those values only that are larger than the mean value, and the other one originating from those values only that are smaller than the mean value. Numerically we find}
\be
0.60 \le \frac{|X_K|}{X(x_t)} \le 1.30\,, \qquad 
0.90 \le \frac{|X_d|}{X(x_t)} \le 1.12\,, \qquad 
0.95 \le \frac{|X_s|}{X(x_t)} \le 1.08\,,
\ee
implying that the CP-conserving effects in the $K$ system can be 
much larger than in the  $B_d$ and $B_s$ systems, where NP effects
are found {to be} disappointingly small.

\begin{figure}
\center{\psfig{file=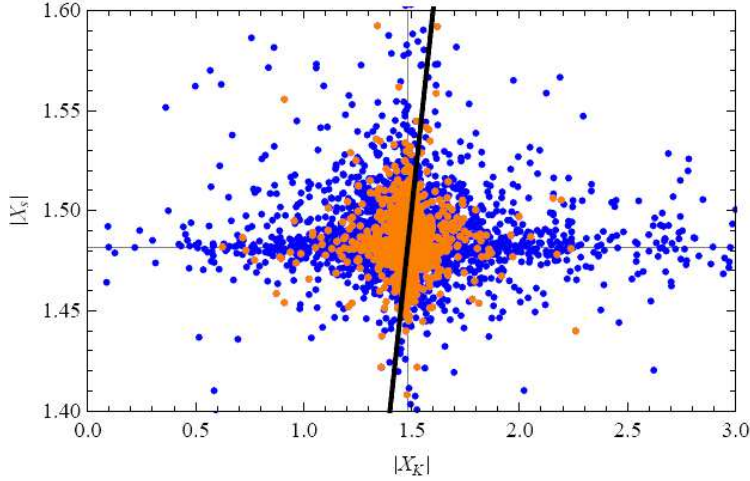,scale=.45}}
\caption{\it Breakdown of the universality between $|X_K|$ and $|X_s|$.
 The solid thick line represents the {CMFV} relation $|X_s|=|X_K|$ and the 
 crossing point of the three solid lines {indicates the SM prediction.}
}
\label{fig:XsXK}
\end{figure}

We illustrate this in Fig.\ \ref{fig:XsXK}, where we show the ranges allowed 
 in the space {$(|X_K|,|X_s|)$}.
The solid thick line represents the {CMFV}
relation $|X_s|=|X_K|$ and the crossing point of the 
thin solid lines {indicates the  SM value}.
The departure from the solid thick line gives the size of {non-CMFV} 
contributions that are caused dominantly by
NP effects in the $K$ system.

Similar hierarchies are  found for 
$|Y_i|$ and $|Z_i|$:
\begin{gather}
0.45 \le \frac{|Y_K|}{Y(x_t)} \le 1.60\,, \qquad 
0.85 \le \frac{|Y_d|}{Y(x_t)} \le 1.20\,, \qquad 
0.93 \le \frac{|Y_s|}{Y(x_t)} \le 1.12\,,\\
0.35 \le \frac{|Z_K|}{Z(x_t)} \le 2.05\,, \qquad 
0.80 \le \frac{|Z_d|}{Z(x_t)} \le 1.30\,, \qquad 
0.90 \le \frac{|Z_s|}{Z(x_t)} \le 1.17\,.
\end{gather}
The fact that largest effects are found in the functions $Z_i$ and the smallest
in the functions $X_i$ {is} dominantly due to the hierarchy
$X(x_t)>Y(x_t)>Z(x_t)$ as
\be
X(x_t)=1.48\,,\qquad  Y(x_t)=0.94\,, \qquad  Z(x_t)=0.65\,.
\ee

\begin{figure}
\center{\psfig{file=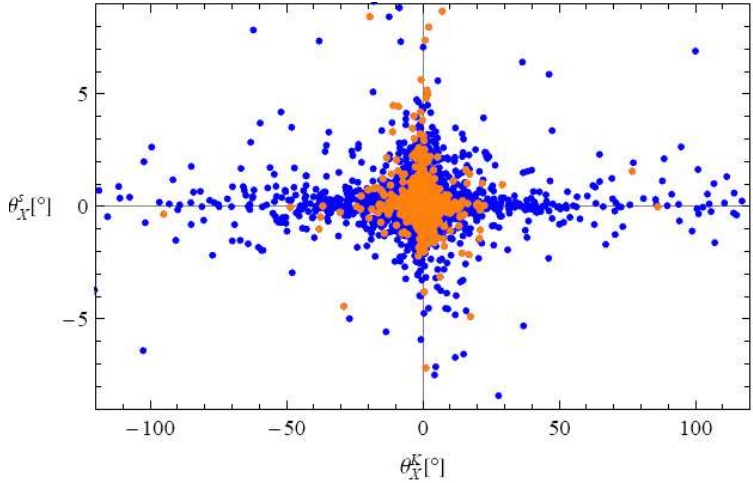,scale=.45}}
\caption{\it Breakdown of the universality between $\theta_X^K$ and $\theta_X^s$ and new sources of CP-violation. {In the SM, $\theta_X^K=\theta_X^s=0$.}}
\label{fig:TsTK}
\end{figure}

For the new complex phases we find  the ranges
\begin{equation}
-45^\circ \le \theta_X^K \le 25^\circ\,, \qquad 
-9^\circ \le \theta_X^d\le 8^\circ\,,\qquad
-2^\circ \le \theta_X^s\le 7^\circ\,,
\label{eq:ThetaX}
\end{equation}
implying that the new CP-violating effects in the {$b\to d\nu\bar\nu$ and $b\to s\nu\bar\nu$ transitions}
are very small, while those in $K_L$ decays can be sizable.
An analogous pattern is found for the phases of $Y_i$ and $Z_i$ functions:
\begin{gather}
-60^\circ \le \theta_Y^K \le 38^\circ\,, \qquad 
-15^\circ \le \theta_Y^d\le 12^\circ\,,\qquad
-4^\circ \le \theta_Y^s\le 11^\circ\,,\label{eq:ThetaY}\\
\qquad-80^\circ \le \theta_Z^K \le 55^\circ\,, \qquad 
-21^\circ \le \theta_Z^d\le 17^\circ\,,\qquad
-6^\circ \le \theta_Z^s\le 15^\circ\,.\label{eq:ThetaZ}
\end{gather}
Again the largest effects are found in the $Z_i$ functions.
As an example we show
in Fig.\ \ref{fig:TsTK} the allowed range in the space
{$(\theta_X^K,\theta_X^s)$}. 

From these results it is evident that {flavour universality can be  significantly violated}.
The anatomy of the hierarchies in the factors
 $1/\lambda_t^{(i)}$ and in the gauge couplings of $Z$ leading to this
breakdown and to its particular pattern can be found in Section~\ref{sec:anatomy}.

\boldmath
\subsection{The $K\to \pi\nu\bar\nu$ System}
\unboldmath

\begin{figure}
\begin{minipage}{8.2cm}
{\psfig{file=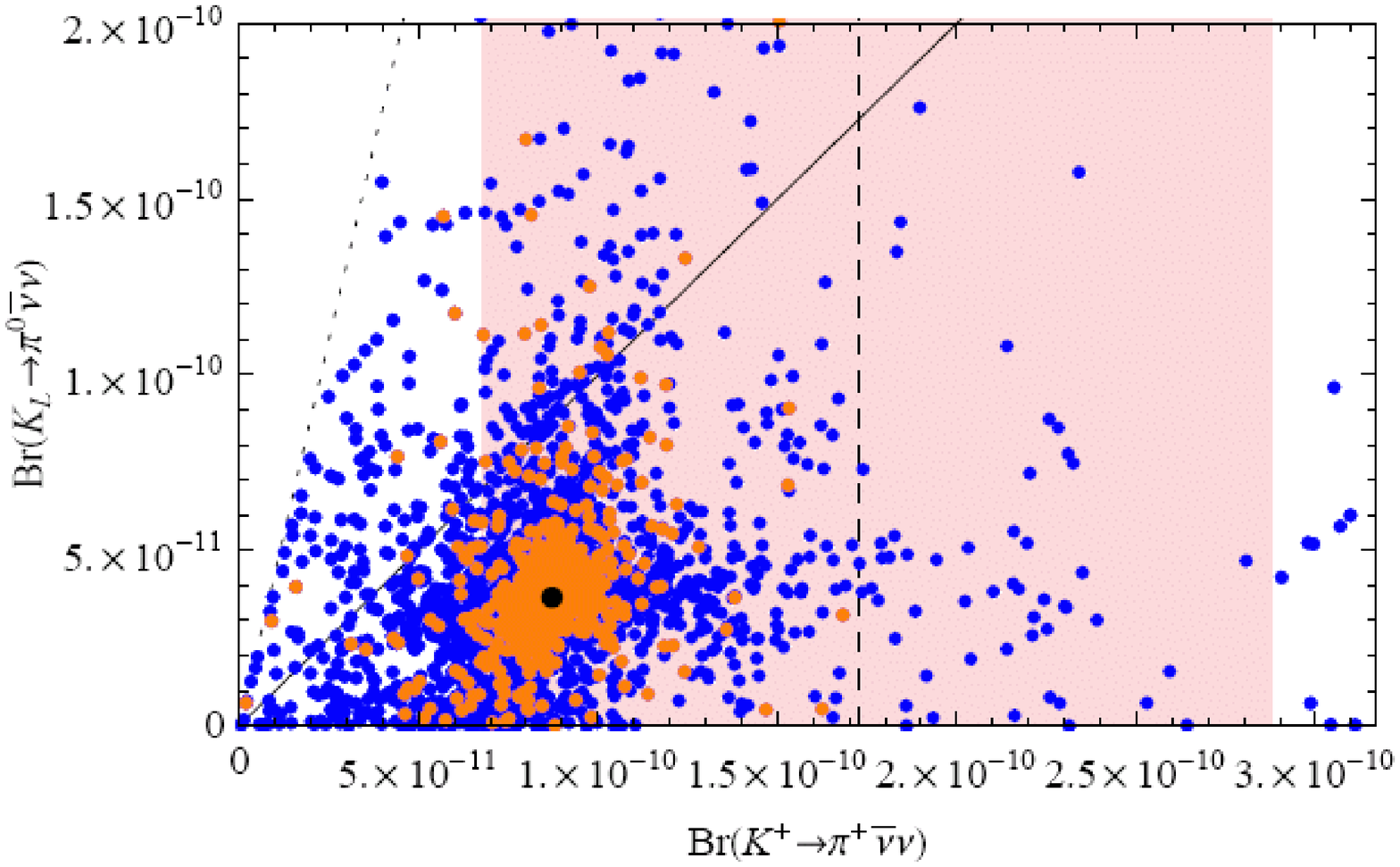,scale=.32}}
\end{minipage}
\begin{minipage}{7cm}
\center{\psfig{file=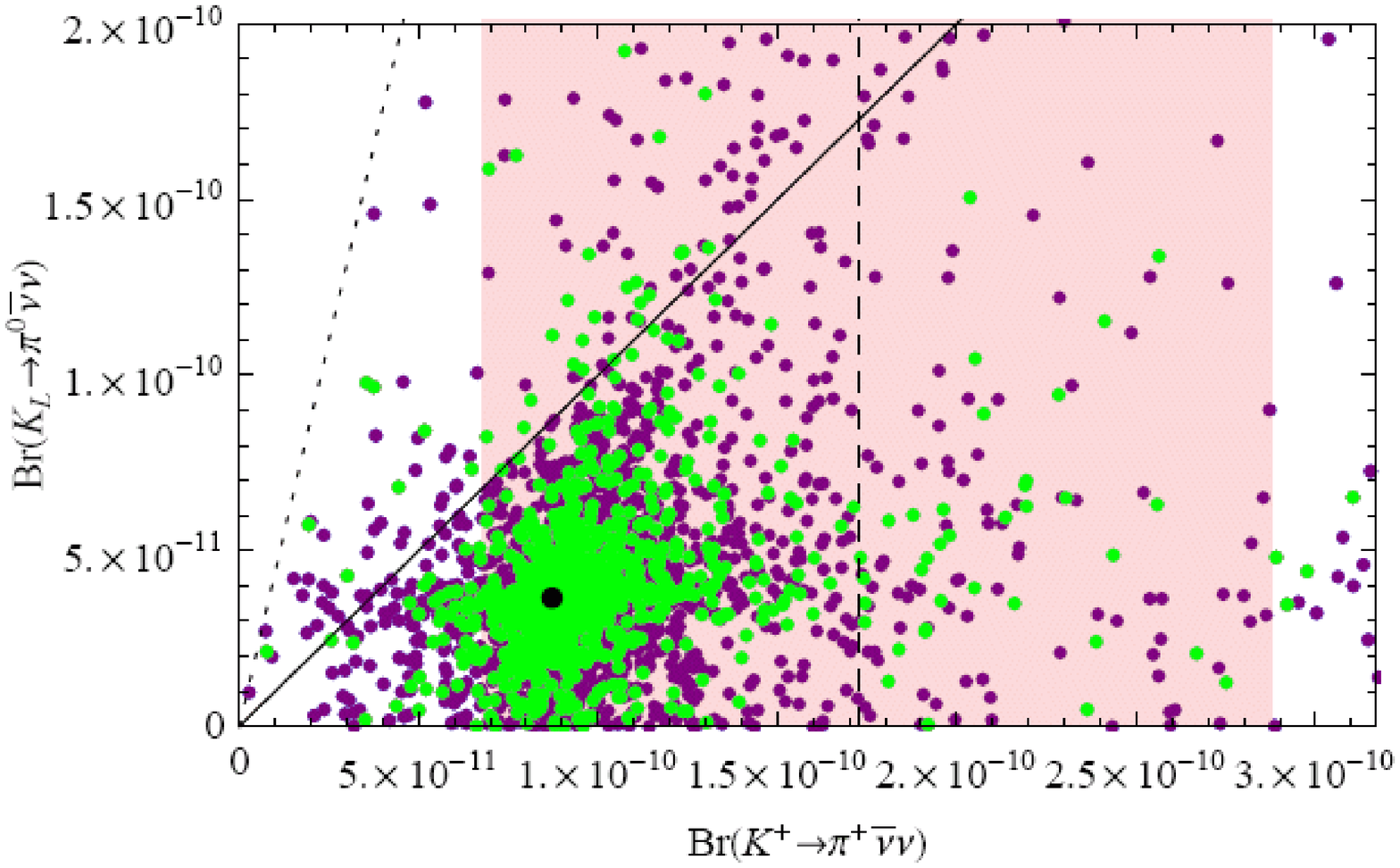,scale=.32}}
\end{minipage}
\caption{\it Left: $Br(\klpn)$ as a function of $Br(\kpn)$. The shaded
    area represents the experimental $1\sigma$-range for $Br(\kpn)$. The
    GN-bound is displayed by the dotted line, while the solid line
    separates the two areas where $Br(\klpn)$ is larger or smaller than
    $Br(\kpn)$. {The black point represents the SM prediction.}
{Right: The same, but in the case of removed custodial protection.}}
\label{fig:KLKp}
\end{figure}

In {the left panel of} Fig.\ \ref{fig:KLKp} we show the correlation between $Br(\kpn)$ and 
$Br(\klpn)$. The experimental
$1\sigma$-range for $Br(\kpn)$ \cite{Artamonov:2008qb} and the
model-independent Grossman-Nir (GN) bound \cite{Grossman:1997sk} are also
shown. We observe that  
$Br(\klpn)$ can be as large as  $15\cdot 10^{-11}$, that is by a factor of
5 larger than its SM value $(2.8\pm 0.6)\cdot 10^{-11}$ while being still consistent with the measured 
value for $Br(\kpn)$. The latter branching ratio can be enhanced by at
most a factor of 2 but this is sufficient to reach {the} central experimental 
value \cite{Artamonov:2008qb}
\be
Br(\kpn)_\text{exp}=(17.3^{+11.5}_{-10.5})\cdot 10^{-11}\,,
\ee
{to be compared with the SM value \cite{Brod:2008ss}
\be
Br(\kpn)_\text{SM} =(8.5\pm 0.7)\cdot 10^{-11}\,.
\ee}

{
In the right panel of Fig.\ \ref{fig:KLKp} we show the modification  {when the custodial protection for $Z$ couplings is removed} as discussed in Section~\ref{sec:remove}. Now the values of $Br(K_L\rightarrow\pi^0\nu\bar\nu)$ and $Br(K^+\rightarrow\pi^+\nu\bar\nu)$ can be as large as $2\cdot 10^{-10}$ and $3\cdot 10^{-10}$ respectively, i.\,e. in the absence of protection an additional enhancement by almost a factor 2 is possible.}

\boldmath
\subsection{$S_{\psi\phi}$ and $K\to\pi\nu\bar\nu$}
\unboldmath

In our previous paper 
\cite{Blanke:2008zb}  spectacular NP effects in
{the CP-asymmetries $S_{\psi\phi}$ and  $A^s_{\rm SL}$} have been found.
{Therefore let us now} have a closer look
at the correlations between $S_{\psi\phi}$ and the $K\to\pi\nu\bar\nu$ decays.
In Figs.~\ref{fig:KLS} and~\ref{fig:KpS} we show the correlation
 between $S_{\psi\phi}$ and 
$Br(\klpn)$ and $Br(\kpn)$, respectively. We observe that
 it is very
difficult to obtain simultaneously large deviations from the SM
 in  the $K\to \pi \nu\bar\nu$ decays and in $S_{\psi\phi}$.

\begin{figure}
\center{\psfig{file=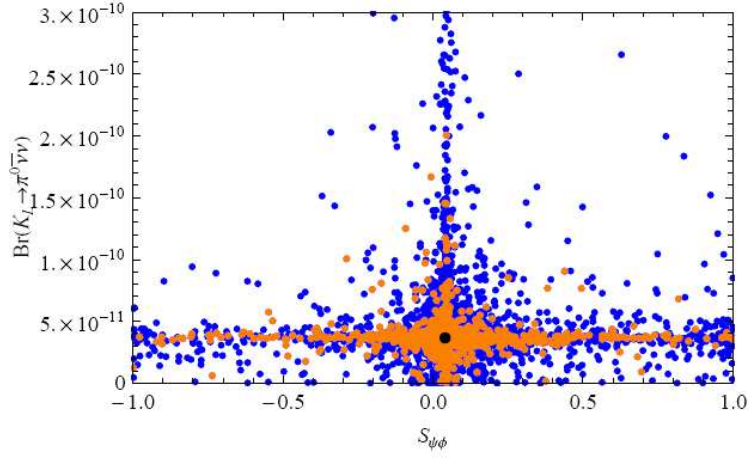,scale=.45}}
\caption{\it $Br(K_L \to \pi^0
\nu \bar \nu)$ as a
  function of $S_{\psi \phi}$. {The black point represents the SM prediction.}}
  \label{fig:KLS}
\end{figure}

\begin{figure}
\center{\psfig{file=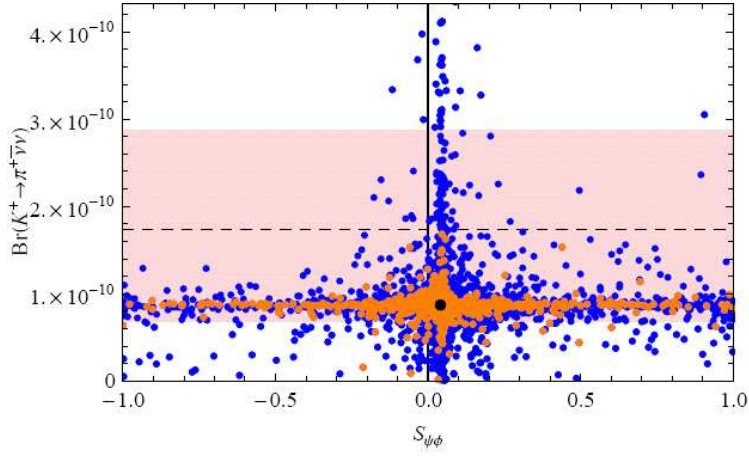,scale=.45}}
\caption{\it $Br(K^+ \to \pi^+
\nu \bar \nu)$ as a
  function of $S_{\psi \phi}$. The shaded area represents the experimental 
             $1\sigma$-range for 
             $Br(\kpn)$, {and the black point the SM prediction.}}
\label{fig:KpS}
\end{figure}

\boldmath
\subsection{$B\to K\nu\bar\nu$, $B\to K^*\nu\bar\nu$ and 
$B \to X_{s,d}\nu\bar\nu$}
\unboldmath

\begin{figure}
\center{\psfig{file=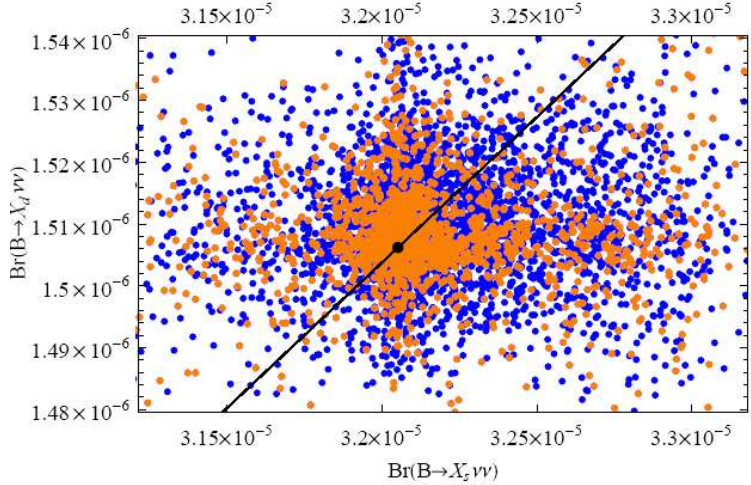,scale=.45}}
\caption{\it Correlation between
$Br(B\to X_s\nu\bar\nu)$ and $Br(B\to X_d\nu\bar\nu)$. The black line
represents {the} universal CMFV result given by the ratio $\vtd^2/\vts^2$, {and the black point the SM prediction.}
 }
\label{fig:bsbd}
\end{figure}

Using the formulae of Sections \ref{sec:excl} and \ref{sec:incl} we find
the ranges
\begin{gather}
0.90\le R_1=\frac{Br(B^+\to K^+\nu\bar\nu)}{Br(B^+\to K^+\nu\bar\nu)_{\rm SM}}\le 1.15\,, \qquad   0.90\le R_2 \le 1.10 \,,\\
0.95\le\frac{Br(B\to X_s\nu\bar\nu)}{Br(B\to X_s\nu\bar\nu)_{\rm SM}}
\le 1.08\,,
\end{gather}
and
\be
0.93\le P \le 1.07\,.
\ee
These results show that NP effects in rare $B$ decays are significantly
smaller than in rare $K$ decays as already expected from our anatomy of
NP effects in Sections \ref{sec:anatomy} and \ref{subsec:UB}.
 As the deviation of $P$ from unity signals violation of an
important and very clean correlation between  
$Br(B\to X_s\nu\bar\nu)$ and $Br(B\to X_d\nu\bar\nu)$ in the CMFV models
 we show this correlation
in Fig.\ \ref{fig:bsbd}. {Unfortunately, the resulting deviation is small
and will be difficult to measure. Therefore we do not show 
 the correlation in \eqref{eq:rel-nn}  that would display strong deviations
 from CMFV mainly due to large effects in $\klpn$ but small ones in
 $B\to X_s\nu\bar\nu$. Similar effects have already been seen in several plots in our paper in the  case of other correlations between $K$ and $B$ decays.}

\boldmath
\subsection{$B_{s}\to \mu^+\mu^-$ versus $K^+\to\pi^+\nu\bar\nu$}
\unboldmath

\begin{figure}
\begin{minipage}{8.2cm}
{\psfig{file=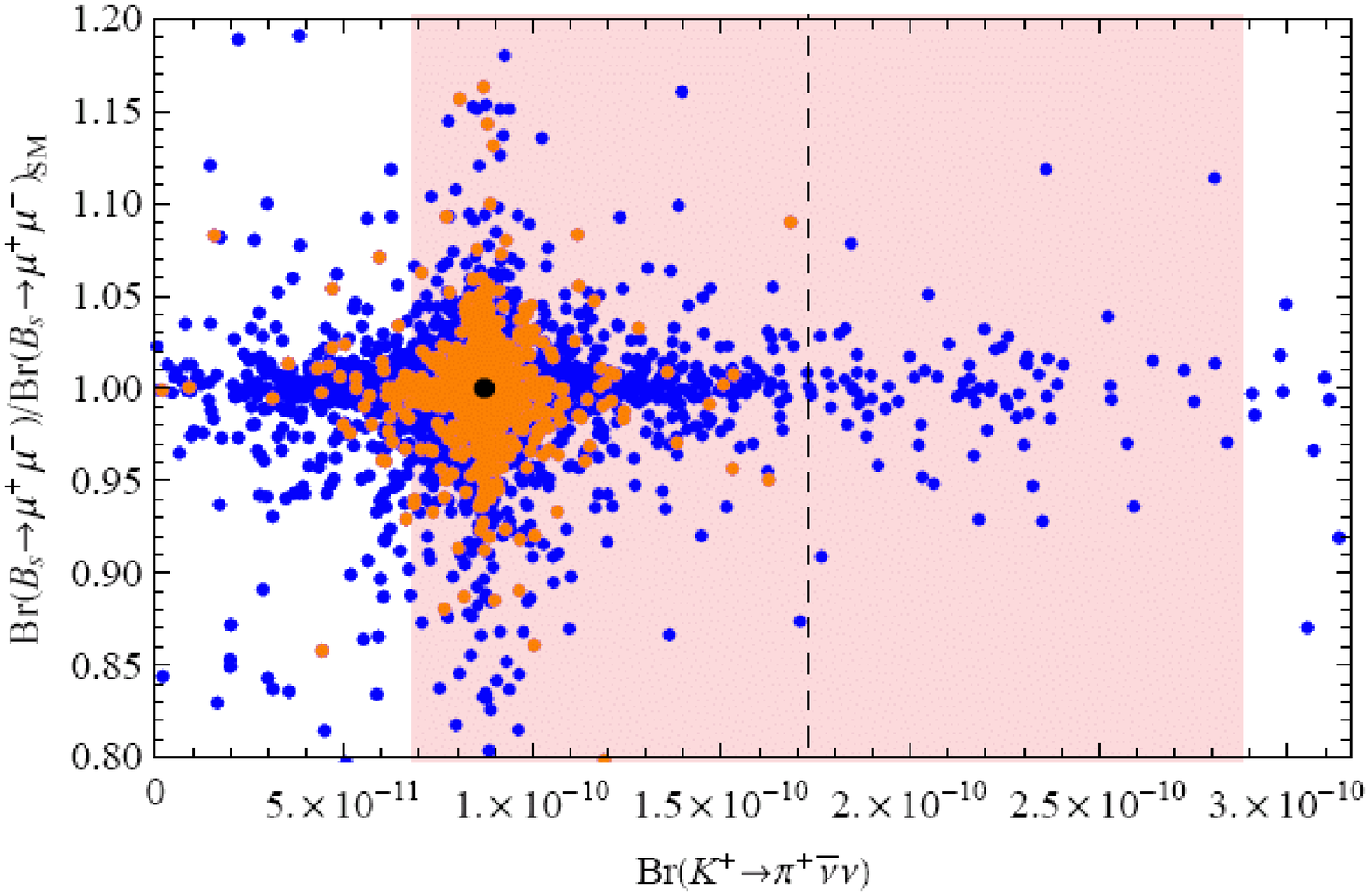,scale=.32}}
\end{minipage}
\begin{minipage}{7cm}
\center{\psfig{file=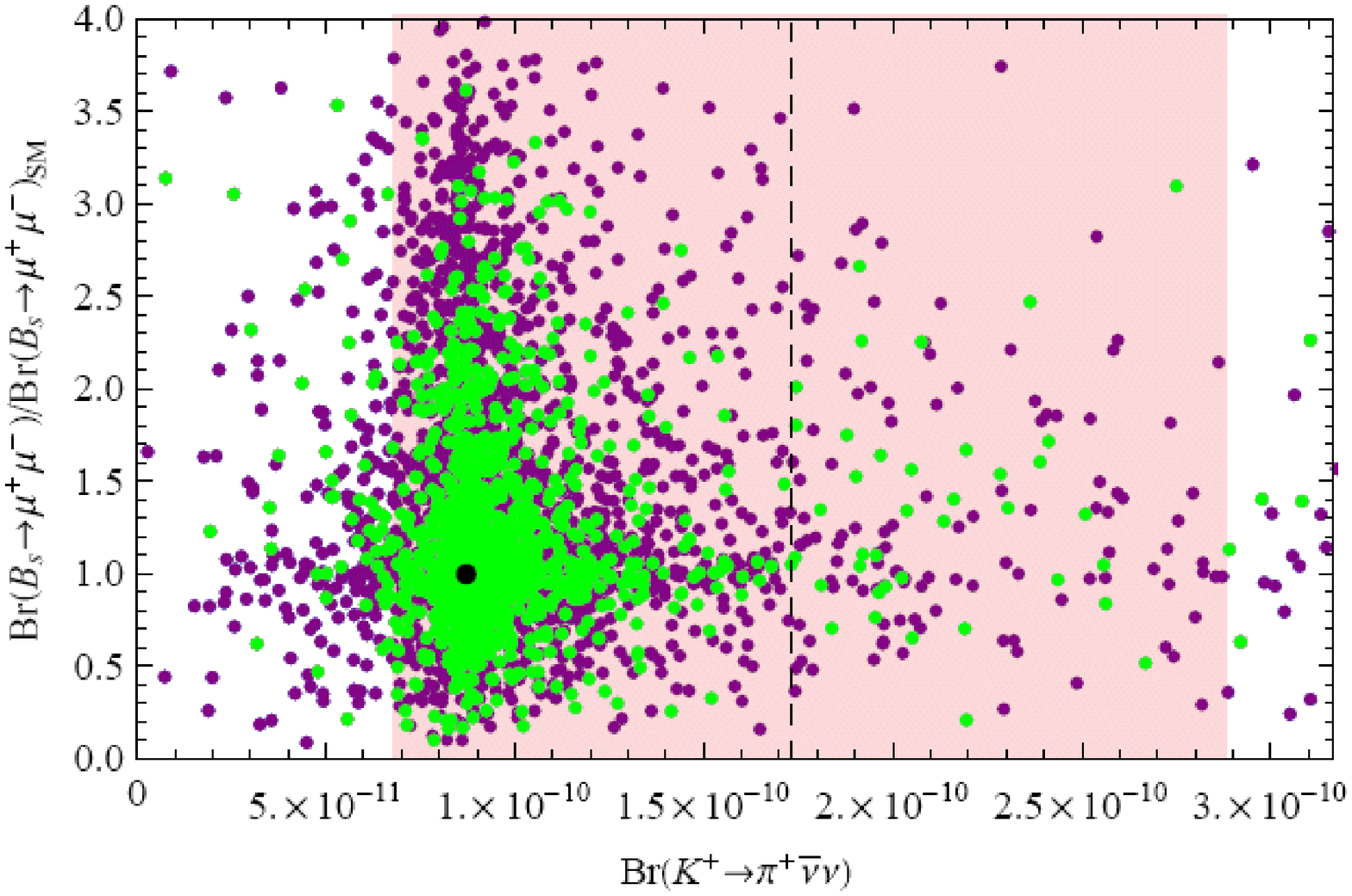,scale=.32}}
\end{minipage}
\caption{\it Left: $Br(B_s\to \mu^+\mu^-)/Br(B_s\to \mu^+\mu^-)_\text{SM}$ as a
  function of $Br(\kpn)$. The shaded area represents the experimental
  $1\sigma$-range for $Br(\kpn)$ and the black point shows the SM prediction.
{Right:  The same, but in the case of removed custodial protection.}}
\label{fig:BsKp}
\end{figure}

We next investigate {the} possible correlation between
$B_{s}\to \mu^+\mu^-$ and $K^+\to\pi^+\nu\bar\nu$. To this end we
show in {the left panel of} Fig.\ \ref{fig:BsKp} the correlation between
$Br(B_s\to \mu^+\mu^-)/Br(B_s\to \mu^+\mu^-)_\text{SM}$ {and}
 $Br(\kpn)$.
The experimental
$1\sigma$-range for $Br(\kpn)$ \cite{Artamonov:2008qb} is represented by the shaded area
and the SM prediction by the black point.
$Br(B_s\to \mu^+\mu^-)$ can deviate only by  $15\%$ from the SM value, while
more pronounced effects are possible in $Br(\kpn)$ as we have seen
before. Again, when $Br(\kpn)$ is sizably enhanced, $Br(B_s\to \mu^+\mu^-)$
{can hardly} be distinguished from the SM prediction.

{
The situation changes spectacularly\footnote{Similarly spectacular effects
of the removal of protection are found in $B\to X_{s,d}\nu\bar\nu$, as opposed to tiny effects in Fig.\ \ref{fig:bsbd}.} when the protection of $Z$ couplings is removed, as shown in the right panel of Fig.\ \ref{fig:BsKp}.
 While $Br(K^+\rightarrow\pi^+\nu\bar\nu)$ can now be enhanced by another factor {of two}, a much bigger effect is seen in the case of $Br(B_s\rightarrow\mu^+\mu^-)$. 
More precisely, the possible effects are now roughly of equal size in both decays, with even  slightly bigger effects observed in $B_s\to\mu^+\mu^-$. This pattern can be easily understood from the discussion in Section \ref{sec:remove}: In the absence of custodial protection the NP effects are clearly dominated by $\Delta_L^{ij}(Z)$ which exhibits a similar hierarchy as the relevant CKM factors $\lambda_t^{(q)}$. The slightly bigger effects  in $B_s\to\mu^+\mu^-$ are then a remnant of the hierarchy $X(x_t)>Y(x_t)$ in the SM, implying that NP effects are generally more pronounced in the latter case.
We would like to note however that such large enhancements in $B_s\to\mu^+\mu^-$ are generally expected to coincide with a violation of the $Z b_L \bar b_L$ constraint, so that a more thorough analysis including also this latter constraint is required to make a definite prediction in the model without custodial protection.
}


\boldmath
\subsection{Correlation between $K_L\to\mu^+\mu^-$ and $B_s\to\mu^+\mu^-$}
\unboldmath

\begin{figure}
\begin{minipage}{8.2cm}
{\psfig{file=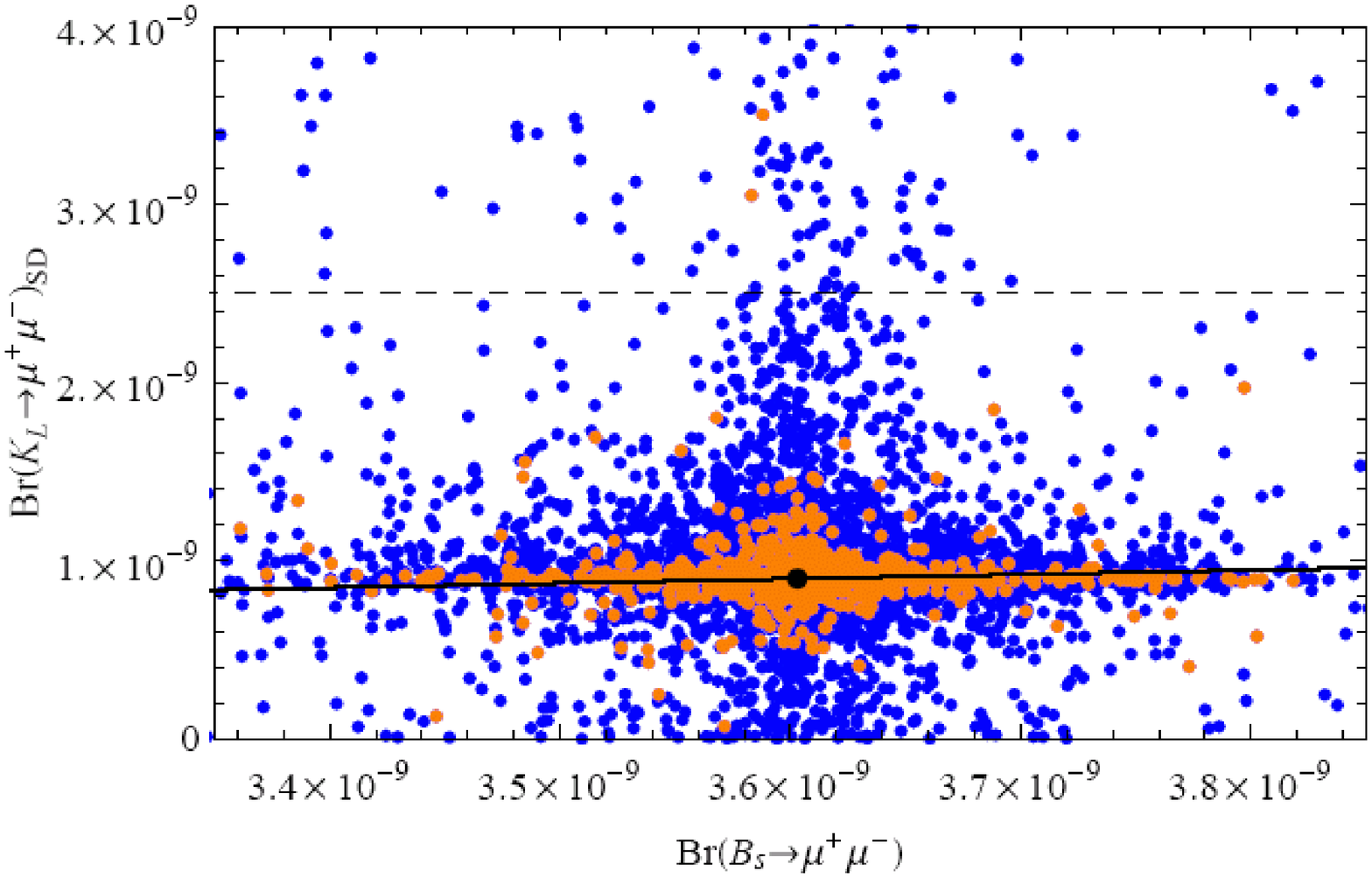,scale=.32}}
\end{minipage}
\begin{minipage}{7cm}
\center{\psfig{file=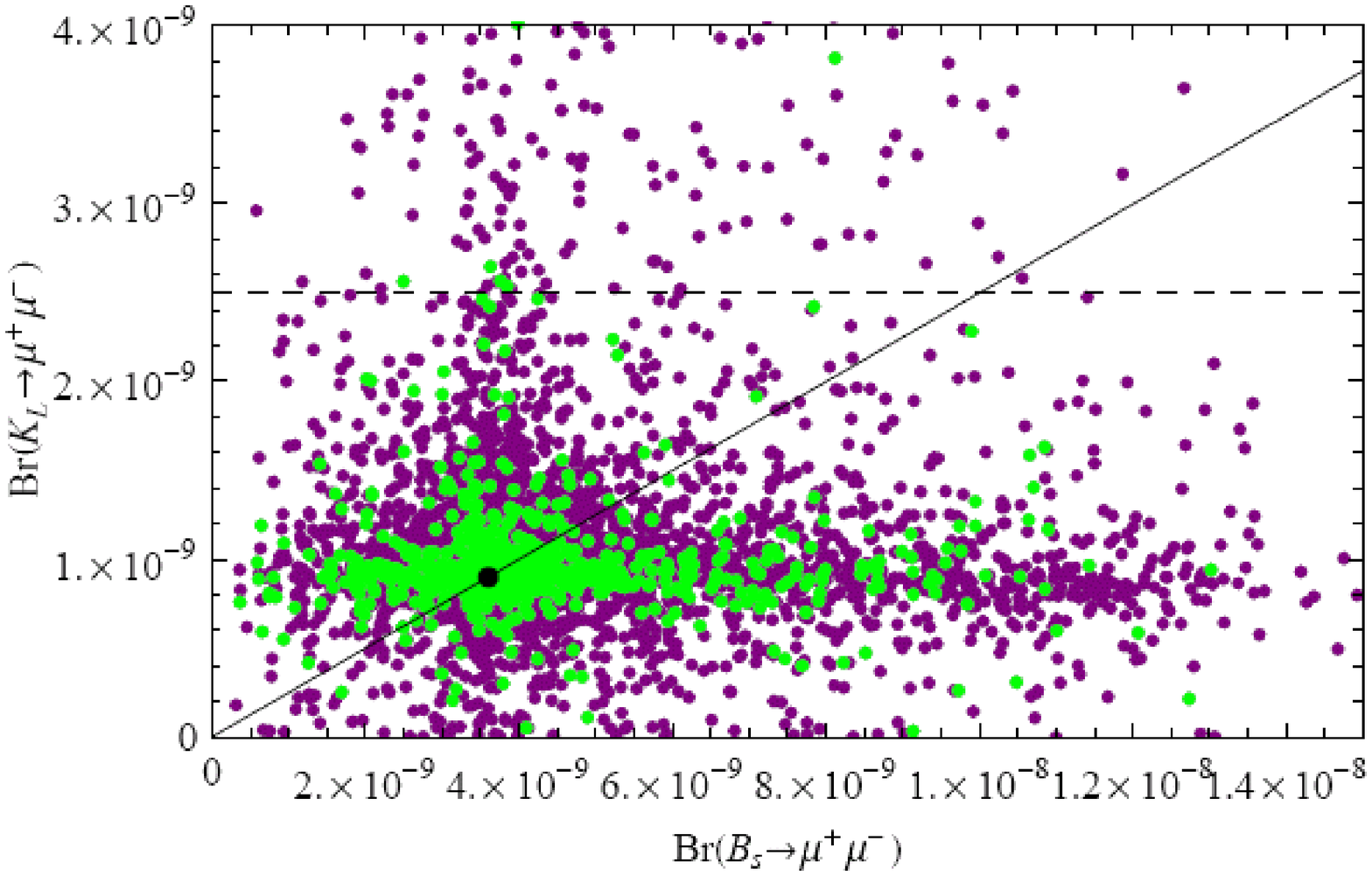,scale=.32}}
\end{minipage}
\caption{\it left: $Br(K_L\to\mu^+\mu^-)_\text{SD}$ as a
  function of $Br(B_s\to\mu^+\mu^-)$. The dashed line indicates the upper bound on $Br(K_L\to\mu^+\mu^-)_\text{SD}$. {The solid line shows the CMFV prediction, while the black point represents the SM.}
{right:  The same, but in the case of removed custodial protection.}}
\label{fig:16}
\end{figure}

 In {the left panel of} Fig.\ \ref{fig:16} we show the correlation between the branching ratios
 for $K_L\to\mu^+\mu^-$ and $B_s\to\mu^+\mu^-$. As expected from our anatomy
 of NP effects, the CMFV correlation represented by the solid line
 is strongly broken, mainly due to much larger NP effects in the decay of 
 $K_L$. We also observe that the upper bound of (\ref{eq:KLmm-bound}) can be saturated,
 but this happens only for {SM-like} values of $Br(B_s\to\mu^+\mu^-)$.
 
 In {the right panel of} Fig.\ \ref{fig:16} we show the {same} correlation  in the absence of 
custodial protection for the $Z$ couplings to left-handed {down-type} quarks. {The NP
effects are now significantly larger, in particular in $B_s\to\mu^+\mu^-$, as already observed previously. Also in $K_L\to\mu^+\mu^-$ an additional enhancement by a factor {of two} is possible, so that the bound of (\ref{eq:KLmm-bound}) can be 
strongly violated. Interestingly, while the possible effects in $K_L\to\mu^+\mu^-$ and $B_s\to\mu^+\mu^-$ are now very similar, they are generally not expected to appear simultaneously, so that a strong violation of the CMFV prediction displayed by the solid line is possible.
}

{

\boldmath
\subsection{Correlation between $K_L\to \mu^+\mu^-$ and $\kpn$}
\unboldmath

\begin{figure}
\center{\psfig{file=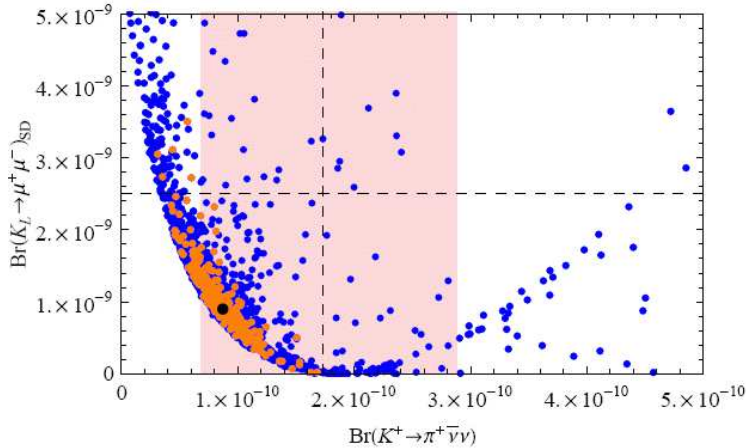,scale=.45}}
\caption{\it Correlation between $K_L\to \mu^+\mu^-$ and $\kpn$. {The black point represents the SM prediction.}}
\label{fig:new2}
\end{figure}

Next in Fig.\ \ref{fig:new2} we show the correlation between the short distance contribution to $Br(K_L\to \mu^+\mu^-)$ and $Br(\kpn)$. As both are CP-conserving rare $K$ decays, a non-trivial correlation is generally expected. Interestingly it turns out that this correlation is an inverse one, i.\,e. an enhancement of $Br(K_L\to \mu^+\mu^-)$ relative to the SM coincides with a suppression of $Br(\kpn)$ and vice versa. 
This correlation originates in the fact that the $K^+\to\pi^+\nu\bar\nu$ transition is sensitive to the vector component of the flavour violating $Z$ coupling, while the $K_L\to\mu^+\mu^-$ decay measures its axial component. As the SM flavour changing $Z$ penguin is purely left-handed, while the NP contribution is dominated by right-handed $Z$ couplings, these two contributions enter the decays in question with opposite relative sign. In other words, the correlation between $K^+\to\pi^+\nu\bar\nu$ and $K_L\to\mu^+\mu^-$ offers a clear test of the handedness of NP flavour violating interactions.

}

\boldmath
\subsection{The $K_L\to \pi^0\ell^+\ell^-$  System}
\unboldmath
\begin{figure}
\center{\psfig{file=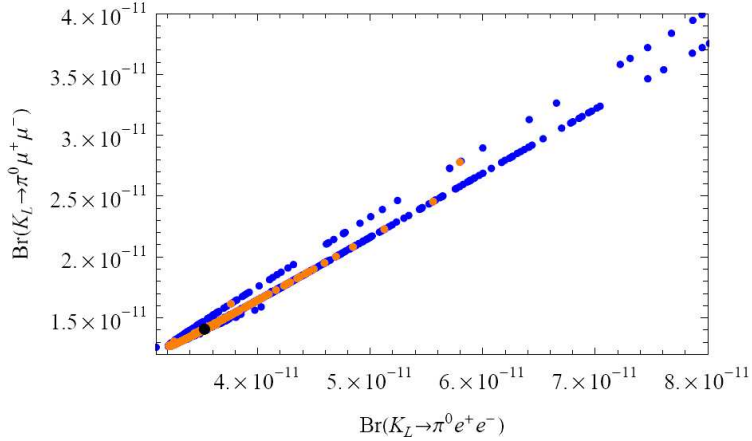,scale=.3}}
\caption{\it $Br(K_L \to \pi^0 \mu^+\mu^-)$ as a
  function of $Br(K_L\to \pi^0 e^+e^-)$, {assuming constructive interference.} {The black point represents the SM prediction.}}
\label{fig:KmuKe}
\end{figure}
In Fig.\ \ref{fig:KmuKe} we show the correlation between 
$Br(K_L\to \pi^0 e^+e^-)$ and $Br(K_L \to \pi^0 \mu^+\mu^-)$
that has {first been} investigated in 
\cite{Isidori:2004rb,Friot:2004yr,Mescia:2006jd}.
We observe that both branching ratios can be enhanced at most by $60\%$
over
the SM values
\cite{Mescia:2006jd}
\begin{gather}
Br(K_L\to\pi^0e^+e^-)_\text{SM}=
3.54^{+0.98}_{-0.85}\left(1.56^{+0.62}_{-0.49}\right)\cdot 10^{-11}\,,\label{eq:KLpee}\\
Br(K_L\to\pi^0\mu^+\mu^-)_\text{SM}= 1.41^{+0.28}_{-0.26}\left(0.95^{+0.22}_{-0.21}\right)\cdot 10^{-11}\label{eq:KLpmm}\,,
\end{gather}
with the values in parentheses corresponding to the destructive interference
between directly and indirectly CP-violating contributions. 
{A recent discussion  of the theoretical status of this interference
sign can be found in \cite{Prades:2007ud} where the results of \cite{Isidori:2004rb,Friot:2004yr,Bruno:1992za} are
critically analysed. From this discussion, constructive interference
seems to be  favoured though more work is necessary.\footnote{We thank  Joaquim Prades for clarifying comments.}}
In Fig.\ \ref{fig:KmuKe} {constructive} interference has been
assumed. 
{We also observe a strong correlation between $Br(K_L\to \pi^0 e^+e^-)$ and $Br(K_L \to \pi^0 \mu^+\mu^-)$, {similar} to the case of the LHT model. Indeed such a correlation is common to all models with no scalar operators contributing to the decays in question \cite{Isidori:2004rb,Friot:2004yr,Mescia:2006jd}.}

The present experimental bounds
\be
Br(K_L\to\pi^0e^+e^-)_\text{exp} <28\cdot10^{-11}\quad\text{\cite{AlaviHarati:2003mr}}\,,\qquad
Br(K_L\to\pi^0\mu^+\mu^-)_\text{exp} <38\cdot10^{-11}\quad\text{\cite{AlaviHarati:2000hs}}\,,
\ee
are still by one order of magnitude larger than the SM predictions.

\boldmath
\subsection{$K_L\to \pi^0\ell^+\ell^-$  versus $\klpn$}
\unboldmath
\begin{figure}
\center{\psfig{file=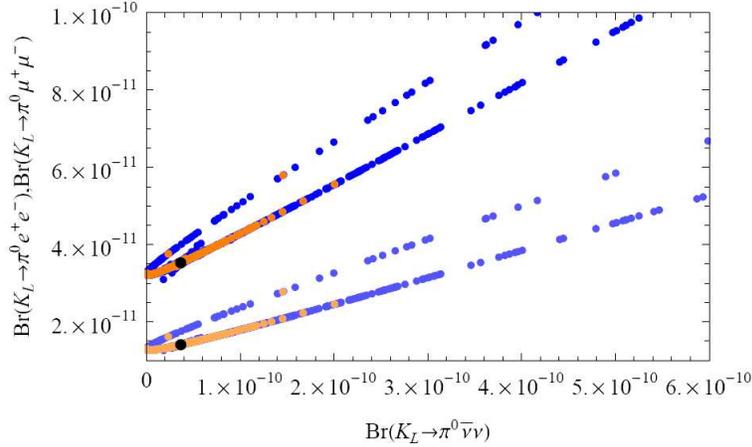,scale=.3}}
\caption{\it $Br(K_L\to \pi^0 e^+e^-)$ (upper curve) and  $Br(K_L \to \pi^0
  \mu^+\mu^-)$ (lower curve) as functions of $Br(\klpn)$. The corresponding SM
  predictions are represented by black points. 
}
\label{fig:KemuKp}
\end{figure}
In Fig.\ \ref{fig:KemuKp}  we show
 $Br(K_L\to\pi^0 e^+e^-)$ and 
$Br(K_L\to\pi^0 \mu^+\mu^-)$ versus $Br(\klpn)$. 
We observe a strong correlation between the $K_L\to \pi^0\ell^+\ell^-$ and 
$\klpn$ decays that has already been found in the LHT model 
\cite{Blanke:2006eb}.
 We note that
a large enhancement of $Br(\klpn)$ automatically implies significant 
enhancements of $Br(K_L\to \pi^0\ell^+\ell^-)$, although the NP effects in
$\klpn$ are much stronger. This is related to the fact that NP effects in 
$K_L\to \pi^0\ell^+\ell^-$ are shadowed by the dominant indirectly
 CP-violating contribution. The correlations in  Figs.~~\ref{fig:KmuKe} and
 \ref{fig:KemuKp} constitute a powerful test of the model considered.
{Again the correlation observed here is very similar to the one encountered in the LHT model \cite{Blanke:2006eb}.}

\subsection{Violation of Golden MFV Relations}\label{sec:golden}

\begin{figure}
\center{\psfig{file=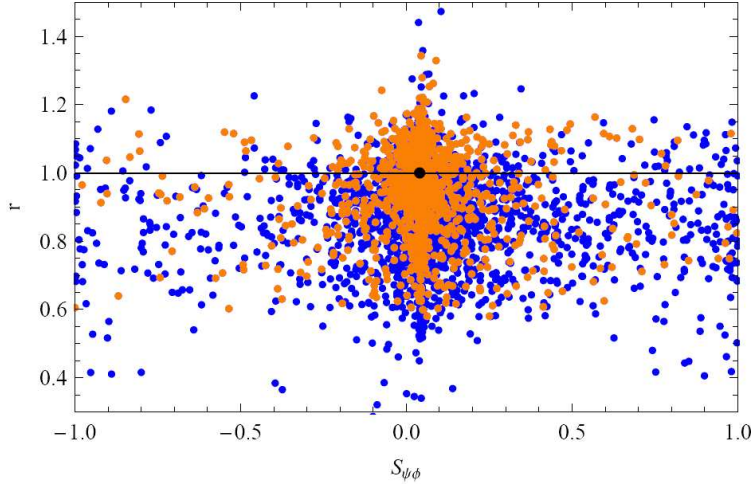,,scale=.3}}
\caption{\it The ratio $r$ {of (\ref{eq:r})} as a function of $S_{\psi\phi}$. {The solid line indicates the CMFV prediction and the black point the SM value.}}
\label{fig:rd13d}
\end{figure}
There are two {\it golden} relations that are theoretically very clean and 
consequently are very suitable for the tests of the SM and its extentions.

We have first
the  relation between $Br(B_{d,s}\to\mu^+\mu^-)$ and $\Delta
M_d/\Delta M_s$ valid in CMFV models \cite{Buras:2003td} that in the model
in question and also in the LHT model gets modified as
follows:
 \be\label{eq:r}
\frac{Br(B_s\to\mu^+\mu^-)}{Br(B_d\to\mu^+\mu^-)}= \frac{\hat
B_{B_d}}{\hat B_{B_s}} \frac{\tau(B_s)}{\tau(B_d)} \frac{\Delta
M_s}{\Delta M_d}\,r\,,\quad r= \left|\frac{Y_s}{Y_d}\right|^2
\frac{C_{B_d}}{C_{B_s}}\,,\qquad 
{C_{B_{d,s}}= \frac{\Delta M_{d,s}}{(\Delta M_{d,s})_\text{SM}}\,,}
\ee
with $r=1$ in {CMFV} models but generally different from unity.

In Fig.\ \ref{fig:rd13d} we show the ratio $r$ of (\ref{eq:r}) as a
function of $S_{\psi\phi}$.  The
departure from unity measures the violation of the golden {CMFV} relation
between $B_{d,s}\to\mu^+\mu^-$ decays and $\Delta M_{d,s}$ in (\ref{eq:r}).
We observe that $r$ can vary roughly in the range
\be
{0.60} \le r \le 1.35\,,
\ee
{with only a weak correlation with $S_{\psi\phi}$.}

\begin{figure}
\center{\psfig{file=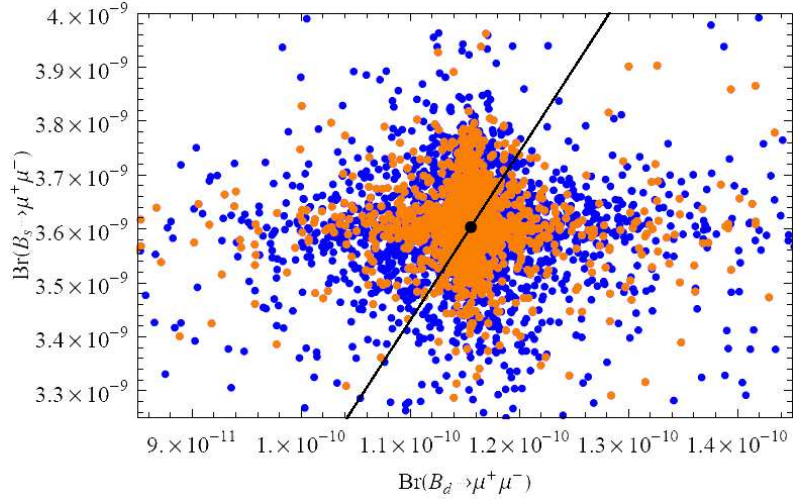,scale=.38}}
\caption{\it 
$Br(B_s\to\mu^+\mu^-)$ versus $Br(B_d\to\mu^+\mu^-)$. The straight line 
represents the CMFV {correlation} {and the black point the SM prediction.}
}
\label{fig:new}
\end{figure}

It is instructive to show also
the plot of 
$Br(B_s\to\mu^+\mu^-)$ versus $Br(B_d\to\mu^+\mu^-)$ that in CMFV models
is a straight line with the slope given in (\ref{eq:r}) with $r=1$. {A similar strong correlation within general MFV models exists \cite{Hurth:2008jc}.}
As shown in Fig.\ \ref{fig:new} deviations from this straight line signal 
non-{CMFV}
effects present in the model considered.
We note that NP effects in $B_d\to \mu^+\mu^-$ are larger than in
$B_s\to \mu^+\mu^-$ as expected from our discussion in Section~\ref{subsec:UB}, {but in any case both are small and difficult to be tested in future experiments.}

{Another golden test of the MFV hypothesis is given by the ratio
\be\label{eq:betarel}
\frac{\sin{2\beta_X^K}}{\sin{2(\beta+\varphi_{B_d})}}\,, 
\ee 
i.\,e. by comparing CP-violation in $B_d-\bar B_d$ mixing and the decay $\klpn$. While in models with new flavour and CP-violating interactions, such as the LHT model \cite{Blanke:2006eb,Blanke:2008ac}, this ratio can deviate significantly from unity, in MFV models $\varphi_{B_d}=0,\,\theta^K_X=0$ holds, so that the MFV relation
of  \cite{Buchalla:1994tr,Buras:2001af}
\be
(\sin2\beta)_{S_{\psi K_S}} = (\sin2\beta)_{\klpn} 
\ee 
is recovered and the ratio in \eqref{eq:betarel} is very close to 1. A violation of this relation would thus clearly
signal the presence of new complex phases and non-MFV
interactions.

\begin{figure}
\center{\psfig{file=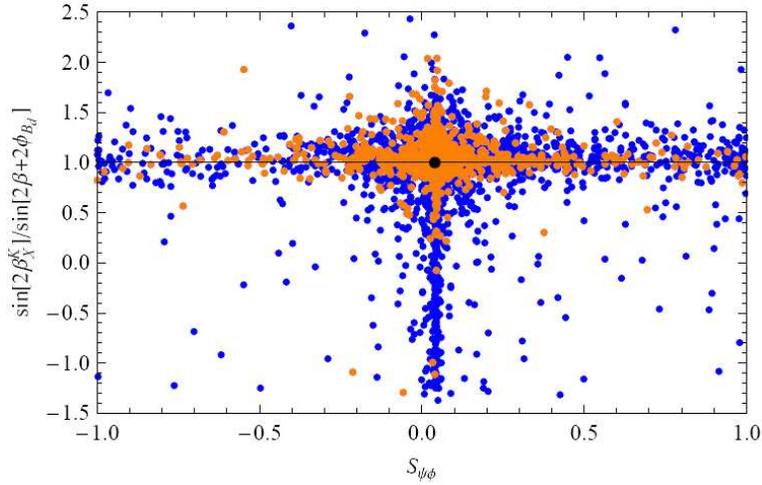,scale=.38}}
\caption{\it $\sin2\beta_X^K/\sin(2\beta+2\varphi_{B_d})$ as a
  function of $S_{\psi\phi}$. {The departure} from unity {(solid line)} measures the size of
  non-MFV effects. {The black point represents the SM prediction.}
\label{fig:rbetad13d}}
\end{figure}
In Fig.\ \ref{fig:rbetad13d} we show the ratio of $\sin 2\beta_X^K$
over $\sin(2\beta+2\varphi_{B_d})$ as a function of {$S_{\psi\phi}$.} 
As discussed in Section~\ref{subsec:5.1},  $\varphi_{B_d}$ is found to
 be small, and
large violations of the relation in question 
can only follow from {large deviations of $X_K$ from its SM value.}
As seen in Fig.\ \ref{fig:rbetad13d}, they can be significant, but again only
for SM-like values of $S_{\psi\phi}$.
}

\boldmath
\subsection{Comparison with the Results in the LHT Model}
\unboldmath
The pattern of deviations from the SM predictions found in the {RS} model
analysed here  and in \cite{Blanke:2008zb} differs from the one found in the
LHT model {\cite{Blanke:2006sb,Blanke:2006eb,Blanke:2008ac}}:
\begin{itemize}
\item
NP effects in $S_{\psi\phi}$ analysed already in \cite{Blanke:2008zb} 
can be large
in both models but the ones in the {RS} model can generally be larger 
{due to the larger number of free flavour parameters and}
to the presence of the $\mathcal{Q}_{LR}$ operators that are absent in the LHT
model.
\item
NP effects in rare $K$ decays can be large in both models but this time
it is easier to enhance the relevant branching ratios in the LHT model, {in particular in the case of  CP-conserving decays like $K^+\to\pi^+\nu\bar\nu$.}
Even if FCNC transitions take place in the {RS} model already at {the} tree level,
the custodial protection of left-handed $Z$ couplings, {the} RS-GIM mechanism
and {masses of the gauge bosons in this model larger {($(2-3)$ TeV)} than the 
masses} of new electroweak gauge bosons in the LHT model (typically smaller
than 1 TeV) taken together do not allow for effects as large as {the} one-loop
effects in the LHT model.
\item
{The correlations between $K_L\to \pi^0\mu^+\mu^-$ and $K_L\to \pi^0e^+e^-$ and between $K_L\to \pi^0\ell^+\ell^-$ and $\klpn$ decays are similar in the RS model considered and in the LHT model.}
\item
{The situation is different for the $K\to\pi\nu\bar\nu$ system, where in the LHT model a strong correlation between $\klpn$ and $K^+\to\pi^+\nu\bar\nu$ has been found \cite{Blanke:2006eb,Blanke:2008ac}. This should be contrasted to the case of the RS model, where we observe no visible correlation between these two decays.
}
\item
{A similar pattern of NP effects is observed in rare $B$ decays but the 
effects are generally smaller than in $K$ decays in both models.}
\item
{A drastically different situation is encountered in the RS model without custodial protection, where the possible NP effects in rare $K$ and $B$ decays are expected to be of equal size. {In particular $Br(B_s\to\mu^+\mu^-)$ can be enhanced by as much as a factor {of three} which is clearly impossible in the LHT model.} However{, as already stated,} {our analysis is not complete, since} in this scenario also a strong violation of the $Z b_L\bar b_L$ constraint is generally predicted and a consistent analysis {should take} into account also EW precision observables.
}

\item
In both models it is unlikely to obtain simultaneously large effects in
$S_{\psi\phi}$ and in rare $K$ decays, but in the  {RS} model considered
here this effect is more pronounced.
\end{itemize}

{In summary, despite the completely different sources of flavour violation in the RS model and in the LHT model, the general pattern of flavour violating observables is similar in both models and makes a distinction non-trivial. Still some signatures would {clearly favour one or the other model.} In particular:
\bi
\item
An observation of the CP-asymmetry $S_{\psi\phi}$ larger than about 0.4 would strongly disfavour the LHT model and favour RS physics.
\item
Another clear {falsification} of the LHT model could be offered by finding the $K\to\pi\nu\bar\nu$ decay rates outside the correlation predicted by the LHT model.
\item
On the other hand, the observation {of} simultaneous large NP effects in $S_{\psi\phi}$ and in rare $K$ decays would put the RS model under severe pressure.
\ei
}

\newsection{Summary and Outlook}\label{sec:summary}

In the present paper we have performed  a detailed analysis of the most
interesting rare decays of $K$ and $B$ mesons
in a warped extra dimensional model with a custodial protection 
of flavour diagonal and flavour {non-diagonal}  $Z$ boson couplings to
left-handed {down-type} quarks. In this model NP contributions come
dominantly   from  tree level exchanges of $Z$ bosons governed by its 
couplings to {right-handed} {down-type} quarks. The contributions  of {the} $Z_H$ boson
are significantly smaller, while those from
 $Z'$ are negligible, being suppressed by
 the custodial protection mechanism, {its} large mass and small couplings to 
leptons. {Also the contributions of the KK photon $A^{(1)}$ are negligible, being suppressed by the small electromagnetic coupling constant and by the electric charge of the down-type quarks.} The anatomy of various contributions has been presented in
Section \ref{sec:anatomy}.

Using the Feynman rules of \cite{Blanke:2008aa}
 we have calculated  the short distance
functions $X_i$, $Y_i$ and $Z_i$ ($i = K, d, s$).
In the  model in question these functions are complex quantities and carry the index
$i$ to signal the breakdown of the universality of FCNC processes, valid in
the  SM {and MFV models}.
The new weak phases in $X_i$, $Y_i$ and $Z_i$, which are absent in the
SM and models with MFV,
imply potential new CP-violating effects beyond the SM and MFV ones.

With the functions $X_i$, $Y_i$ and $Z_i$ at hand, we have 
calculated the branching ratios for a number of interesting rare decays.
In particular, we analysed $K^+ \to \pi^+ \nu \bar \nu$, $K_L
\to \pi^0 \nu \bar \nu$, $K_L\to\pi^0\ell^+\ell^-$, {$K_L\to\mu^+\mu^-$},
 $B_{s,d} \to \mu^+ \mu^-$, $B\to K\nu\bar\nu$, $B\to K^*\nu\bar\nu$
and  $B\to X_{s,d} \nu \bar \nu$.
At all stages of our numerical analysis we took into account the existing
constraints from $\Delta F=2$ processes analysed by us in \cite{Blanke:2008zb}.

The main messages of our paper are as follows:\footnote{All the results quoted here are obtained constraining also the fine-tuning in $\eps_K$, $\Delta_\text{BG}(\eps_K)<20$.}
\begin{itemize}

\item {The most evident departures from the SM predictions are found for
    CP-violating observables that are strongly suppressed within {the SM}. These are the branching ratio for $K_L \to \pi^0 \nu \bar \nu$ and
    the CP-asymmetry $S_{\psi \phi}$ with the latter analysed already in 
\cite{Blanke:2008zb}. $Br\left(K_{L}\rightarrow\pi^0\nu\bar\nu\right)$ can be by a factor of 5 larger than its SM
    value, while $S_{\psi\phi}$ can be enhanced by {more than} an order of
    magnitude. However as clearly seen in Fig.\ \ref{fig:KLS} simultaneous large NP
    effects in both observables are very unlikely.}

\item {The largest departures from SM expectations for $Br\left(K^+\rightarrow\pi^+\nu\bar\nu\right)$ and $Br(K_L \to
  \pi^0 \ell^+ \ell^-)$ amount to factors of $2$ and $1.6$, respectively.
   The enhancement of $Br(\kpn)$ could be welcomed one day if the central
  experimental value will remain in the ballpark of $15\cdot 10^{-11}$ and its
error will decrease. Again, it is very unlikely to get simultaneously large 
NP effects in $\kpn$ and $S_{\psi\phi}$ (Fig.\ \ref{fig:KpS}), while simultaneous large
effects in $\kpn$ and $\klpn$ are possible as clearly seen in Fig.\ \ref{fig:KLKp}.
}

\item{The branching ratios for $B_{s,d} \to \mu^+ \mu^-$ and $B \to X_{s,d}
    \nu \bar \nu$, instead,  are modified by at most $20\%$ and $10\%$, respectively.
    }

\item{Sizable departures from MFV relations between $\Delta M_{s,d}$ and
    $Br(B_{s,d} \to \mu^+ \mu^-)$ and between $S_{\psi K_S}$ and the $K \to
    \pi \nu \bar \nu$ decay rates are possible. This is clearly seen in 
Figs.~\ref{fig:rd13d} and \ref{fig:rbetad13d}.}

\item{The universality of NP effects, characteristic for MFV models,
    can be largely broken, in particular between $K$ and $B_{s,d}$ systems in
     $\Delta F=1$ transitions, where large effects are only
    possible in $K$ decays.}
\item
{The main impact of the extended gauge group on $\Delta F=1$ processes is the suppression of tree level left-handed $Z$ couplings, while the direct contributions of the new gauge bosons play a subdominant role.}
\end{itemize}

In summary, our present analysis of rare $K$ and $B$ decays combined with our
previous analysis of $\Delta F=2$ transitions reveals a clear pattern of NP
effects in FCNC processes predicted by the {RS} model with custodial protection
in question:
\begin{itemize}
\item
{NP effects in $\Delta S=2$ processes are governed by KK gluons, whereas in $\Delta B=2$ transitions the heavy gauge boson $Z_H$ is equally important.}
\item
{NP effects in $\Delta S=1$ transitions are dominated by tree level $Z$ exchanges.}
\item
{Large effects in $S_{\psi\phi}$ are possible.}
\item
{Large effects in $\klpn$ and $\kpn$ {are possible}, even simultaneously.}
\item
{Large enhancements in $K_L\to\mu^+\mu^-$ and $\kpn$ {are possible}, but not simultaneously.}
\item
{Simultaneous large effects in $S_{\psi\phi}$ and in the $K\to\pi\nu\bar\nu$ decays
are very unlikely.}
\item
{NP effects in rare $B$ decays dominated in the SM by {$Z$ penguin} contributions are
generally small and hardly distinguishable from NP effects in MFV models.}
\end{itemize}

This pattern implies that an observation of a large $S_{\psi\phi}$ would in
the context of the model considered here  preclude sizable NP effects in rare
$K$ decays. On the other hand, finding $S_{\psi\phi}$ to be SM-like
will open the road to large NP effects in rare $K$ decays. Independently of the
experimental value of $S_{\psi\phi}$, NP effects in rare $B$ decays are
predicted to be small and an observation of large departures from SM
predictions in future data would put the model considered here in serious difficulties.

Clearly, this pattern of NP effects in FCNC processes originates to a large
extent in the custodial protection of the $Z$ couplings to left-handed {down-type}
quarks. We have shown that removing this protection from the model allows to
obtain significantly larger NP effects in rare $K$ and $B$ decays. {For instance $Br(B_s\to\mu^+\mu^-)$ could be as large as $10^{-8}$.} On the
other hand without this protection it is much harder to obtain an agreement with the electroweak 
precision data for {KK} scales in the reach of {the} LHC.

 Finally, as a byproduct we have presented general formulae for
 effective Hamiltonians including right-handed couplings to gauge
 bosons that can be used in any extension of the SM. {We have also pointed out a number of correlations between various decays (see Section \ref{sec:correlations}) that could turn out to be crucial for distinguishing various NP scenarios.}

\subsubsection*{Acknowledgments}

We thank {Csaba Csaki} {and} Andreas Weiler for useful discussions.
This research was partially supported by {the Graduiertenkolleg GRK 1054, the Deutsche Forschungsgemeinschaft (DFG) under contract BU 706/2-1}, {the DFG Cluster of Excellence `Origin and Structure of the Universe' and by} the German Bundesministerium f{\"u}r Bildung und
Forschung under contract 05HT6WOA. {S.\,G. acknowledges support by the European Community's Marie Curie Research Training Network under contract MRTN-CT-2006-035505 [``HEPTOOLS''].}

\begin{appendix}

\newsection{Couplings of Electroweak Gauge Bosons}\label{app:Deltas}
\boldmath
\subsection{Couplings of $Z$}
\unboldmath

The flavour non-diagonal couplings of $Z$ to down quarks are given by
\begin{equation}\label{A.16}
\Delta_{L,R}^{ij}(Z)= \frac{M_Z^2}{M_{\text{KK}}^2}
\left[- \mathcal{I}^+_1 \Delta_{L,R}^{ij}(Z^{(1)}) +  \mathcal{I}^-_1 \cos\phi\,\cos\psi
\Delta_{L,R}^{ij}(Z^{(1)}_X) 
\right] { + \Delta_{L,R}^{ij}(Z)_\text{KK-fermions}\,,}
\end{equation}
where
\begin{eqnarray}
\mathcal I_1^{+}&=&\frac{1}{L}\int_0^L dy \,e^{-2ky}g(y)h(y)^2 { \simeq\sqrt{2kL}} \,,\\
\mathcal I_1^{-}&=&\frac{1}{L}\int_0^L dy \,e^{-2ky}\tilde g(y)h(y)^2 { \simeq\sqrt{2kL}} \,.
\end{eqnarray}
{Here $h(y)$ is the Higgs profile, with $h(y)\propto \delta(y-L)$ in the present analysis, and $g(y), \tilde g(y)$ are the shape functions of $Z^{(1)}$ and $Z_X^{(1)}$, respectively, differing slightly from each other due to the different boundary conditions on the UV brane.}
$Z^{(1)}$ and $Z^{(1)}_X$ are gauge eigenstates 
and $\Delta_{L,R}^{ij}(Z^{(1)})$ and  $\Delta_{L,R}^{ij}(Z^{(1)}_X) $ 
are the elements of the $3\times 3$ coupling matrices
\be\label{DeltasZ1,ZX1}
\hat\Delta_{L,R}(V)=\mathcal{D}_{L,R}^\dagger\hat\varepsilon_{L,R}(V)\mathcal{D}_{L,R}\qquad (V=Z^{(1)},Z_X^{(1)})\,,
\ee
with $\mathcal{D}_L$ and $\mathcal{D}_R$ being the left- and right-handed down-type flavour mixing matrices, respectively. They have been discussed and calculated in \cite{Blanke:2008zb}. $\hat\varepsilon_{L,R}(Z^{(1)})$ {and} $\hat\varepsilon_{L,R}(Z_X^{(1)})$ are 
diagonal matrices
\be 
\hat\varepsilon_{L,R}(V)=\text{diag} \big(\varepsilon_{L,R}(1)(V), \varepsilon_{L,R}(2)(V), \varepsilon_{L,R}(3)(V)\big)\qquad(V=Z^{(1)},Z_X^{(1)})\,.
\ee

The couplings of $Z^{(1)}$ and $Z_X^{(1)}$ to fermions in the flavour eigenbasis are given by the overlap integrals
\bea\label{eq:B.5}
\varepsilon_L(i)(Z^{(1)})&=&g_{Z,L}^\text{4D}
\frac{1}{L}\int_0^L dy\,e^{ky} \left[f^{(0)}_{L}(y,c_Q^i)\right]^2 g(y)\,,
\\
\varepsilon_R(i)(Z^{(1)})&=&g_{Z,R}^\text{4D}
\frac{1}{L}\int_0^L dy\,e^{ky} \left[f^{(0)}_{R}(y,c_d^i)\right]^2 g(y)\,,
\\
\varepsilon_L(i)(Z_X^{(1)})&=&\kappa_1^\text{4D}
\frac{1}{L}\int_0^L dy\,e^{ky} \left[f^{(0)}_{L}(y,c_Q^i)\right]^2\tilde g(y)\,,
\\
\varepsilon_R(i)(Z_X^{(1)})&=&\kappa_5^\text{4D}
\frac{1}{L}\int_0^L dy\,e^{ky} \left[f^{(0)}_{R}(y,c_d^i)\right]^2 \tilde g(y)\,.
\label{eq:B.8}
\eea
{Further}
\bea
g_{Z,L}^\text{4D}=\frac{g^\text{4D}}{\cos\psi}\left(-\frac{1}{2}+\frac{1}{3}\sin^2\psi\right)\,&,&~~~~\kappa_1^\text{4D}=\frac{g^\text{4D}}{\cos\phi}\left(-\frac{1}{2}-\frac{1}{6}\sin^2\phi\right)\,,\\
g_{Z,R}^\text{4D}=\frac{g^\text{4D}}{\cos\psi}\left(\frac{1}{3}\sin^2\psi\right)\,&,&~~~~\kappa_5^\text{4D}=\frac{g^\text{4D}}{\cos\phi}\left(-1+\frac{1}{3}\sin^2\phi\right)
\,.
\eea

 Here $g^\text{4D}$ is the 4D  $SU(2)_L$ gauge coupling. Moreover $\sin^2\psi\approx\sin^2\theta_W$ and $\sin\phi$, $\cos\phi$ as functions of $\psi$ are given by the formulae
\be
\cos\psi=\frac{1}{\sqrt{1+\sin^2\phi}}\,,\qquad
\sin\psi=\frac{\sin\phi}{\sqrt{1+\sin^2\phi}}
\ee
 and can also be found in \cite{Blanke:2008aa}.

{Finally, 
the contribution to the flavour violating $Z$ couplings originating from the mixing of the fermionic zero modes with their heavy KK partners
turns out to be a sub-leading effect \cite{Blanke:2008zb}. The corresponding formulae are complicated and beyond the scope of this paper. Details will be presented elsewhere. Note that both gauge and fermion KK contributions to $\Delta_{L}^{ij}(Z)$ are suppressed by the custodial protection mechanism.}

The couplings of $Z$ to {$\ell^+\ell^-$} and $\nu\bar \nu$ are standard:
\begin{gather}
{\Delta_L^{\nu\nu}(Z)}= \frac{1}{2}\frac{g^\text{4D}}{\cos\psi}\,,\\
{\Delta_L^{\ell\ell}(Z)}=\frac{g^\text{4D}}{\cos\psi} 
\left(-\frac{1}{2} + \sin^2\psi\right)\,,
\qquad 
{\Delta_R^{\ell\ell}(Z)}=\frac{g^\text{4D}}{\cos\psi}\sin^2\psi\,.
\end{gather}

\boldmath\subsection{Couplings of $Z_H$ and $Z^{\prime}$ to Down Quarks}
\unboldmath
The couplings $\Delta_{L,R}^{ij}(Z_H)$ and $\Delta_{L,R}^{ij}(Z^{\prime})$
with $i,j=d,s,b$ can be conveniently written in terms of the matrices
 $\hat{\Delta}_{L,R}(Z^{(1)})$ and $\hat{\Delta}_{L,R}(Z_X^{(1)})$, given 
above, as follows
\begin{equation}\label{A.1}
\hat{\Delta}_{L,R}(Z_H) = \cos\xi\, \hat{\Delta}_{L,R}(Z^{(1)}) 
+ \sin\xi\, \hat{\Delta}_{L,R}(Z^{(1)}_X)\,,
\end{equation}
\begin{equation}\label{A.2}
\hat{\Delta}_{L,R}(Z^{\prime}) = -\sin\xi\, \hat{\Delta}_{L,R}(Z^{(1)}) 
+ \cos\xi\, \hat{\Delta}_{L,R}(Z^{(1)}_X)\,,
\end{equation}
where $\cos\xi$ and  $\sin\xi$, 
both $\mathcal{O}(1)$, represent the transformation of $Z^{(1)}$ and
$Z^{(1)}_X$ into the mass eigenstates $Z_H$ and $Z^{\prime}$. The
$\mathcal{O}(v^2/M_{\text{KK}}^2)$ terms in this transformation that involve
the $Z$ boson were
 treated separately above. As $\cos\xi$ and $\sin\xi$ 
do not appear in the final expressions in the limit  $M_{Z_H}=M_{Z^{\prime}}=M_{\text{KK}}$,
we do not give explicit formulae for them. They can be found in
\cite{Blanke:2008aa}. {Note that in the limit of exact $P_{LR}$ symmetry
\be
\frac{\cos\xi}{\sin\xi}=\cos\phi \cos\psi\,,
\ee
removing both flavour diagonal and non-diagonal $Z^\prime$ couplings to left-handed down-type quarks. As in the case of $Z'$ the $P_{LR}$ symmetry is violated at the 10\% level, these couplings receive a suppression relative to the $Z_H$ ones by {merely} one order of magnitude.}

\boldmath
\subsection{Couplings of $Z_H$ and $Z^{\prime}$ to Leptons}
\unboldmath

These couplings are defined in an analogous manner to \eqref{DeltasZ1,ZX1} so that we only list the corresponding
replacements:
\begin{enumerate}
\item 
{We neglect all lepton flavour violation effects and set $c_{L,R}=\pm0.7$ universally for all charged leptons and neutrinos. In this limit the flavour mixing matrices corresponding to $\mathcal{D}_{L,R}$ are simply replaced by the identity, and the coupling matrices $\hat \varepsilon_{L,R}$ are proportional to the unit matrix.}
\item In the case of charged leptons we have
\begin{eqnarray}
g_{Z,L}^{\text{4D}}=\frac{g^{\text{4D}}}{\cos\psi} \left(-\frac{1}{2} + \sin^2\psi\right)\,&,&~~~~\kappa_1^{\text{4D}}= -\frac{1}{2} g^{\text{4D}} \cos\phi\,,\\
g_{Z,R}^{\text{4D}}=\frac{g^{\text{4D}}}{\cos\psi}\sin^2\psi\,&,&~~~~\kappa_5^{\text{4D}}= - g^{\text{4D}} \cos\phi\,.
\end{eqnarray}
\item In the case of neutrinos we have
 \begin{eqnarray}
g_{Z,L}^{\text{4D}}= \frac{1}{2}  \frac{g^{\text{4D}}}{\cos\psi}\,&,&~~~~\kappa_1^{\text{4D}}= -\frac{1}{2} g^{\text{4D}} \cos\phi\,,\\
g_{Z,R}^{\text{4D}}= 0\,&,&~~~~\kappa_5^{\text{4D}}= 0\,.
\end{eqnarray}

\end{enumerate}

\boldmath
\subsection{Couplings of $A^{(1)}$}
\unboldmath

In the case of $A^{(1)}$, $\varepsilon_{L,R}(i)$ is given by $(i=1,2,3)$
\be
\varepsilon_{L,R}(i)(A^{(1)}) = {Q_\text{em}} e^\text{4D}\frac{1}{L}\int_0^L dy\,e^{ky} \left[f^{(0)}_{L,R}(y,c_\Psi^i)\right]^2 g(y)\,,
\label{eq:B.1}
\ee
with $g(y)$ being the gauge KK shape function of $A^{(1)}$.

\end{appendix}

\addcontentsline{toc}{chapter}{References}

\providecommand{\href}[2]{#2}\begingroup\raggedright\endgroup

\end{document}